\documentclass[pre, twocolumn, superscriptaddress]{revtex4-1}
\usepackage{amsmath}
\usepackage{amsfonts}
\usepackage{amssymb}
\usepackage{array}
\usepackage{dcolumn}
\usepackage{longtable}
\usepackage{hyperref}

\usepackage{float}
\usepackage{bm}
\usepackage{graphicx}
\usepackage{natbib}
\usepackage{color}
\usepackage{Tomspeak}
\newcommand{\cross}{\times}

\newcommand{\Mop}{{\mathsf M}}
\newcommand{\Kop}{{\mathsf K}}

\newcommand{\Iop}{{\mathsf I}}
\newcommand{\Rop}{{\mathsf R}}
\newcommand{\Mopo}{\Mop_{\Omega E}}
\newcommand{\IdentityMatrix}{{\mathsf 1}}
\newcommand{\xop}[1]{{\mathsf (}\,\vector #1\,{\mathsf )}^\cross}

\begin{document}
\title{Chiral motion in colloidal electrophoresis}
\date{\today}
\author{Lara Braverman}
\email{lbraverman@uchicago.edu}
\author{Aaron Mowitz}
\email{amowitz@uchicago.edu}
\author{Thomas A. Witten}
\email{t-witten@uchicago.edu}
\affiliation{Department of Physics and James Franck Institute, University of Chicago, Chicago, Illinois 60637, United States.}

\begin{abstract}
Asymmetrically charged, nonspherical colloidal particles in general perform complex rotations and oblique motions under an electric field.  The interplay of electrostatic and hydrodynamic forces complicate the prediction of these motions.  We demonstrate a method of calculating the body tensors that dictate translational and rotational velocity vectors arising from an external electric field.  We treat insulating, rigid bodies in the linear-response regime, with indefinitely small electrostatic screening length.  The method represents the body as an assembly of point sources of both hydrodynamic drag and surface electric field.  We demonstrate agreement with predicted electrophoretic mobility to within a few percent for several shapes with uniform and nonuniform charge.  We demonstrate strong chiral twisting motions for colloidal bodies of symmetrical realistic shapes.  The method applies more generally to active colloidal swimmers.

\end{abstract}
\maketitle

    \section{Introduction}\label{sec:introduction}
    An important class of driven-particle motion is swimming; that is, propulsion through a fluid without external forces on the particles.  Swimming motion can be driven by chemical reactions at the surface of the particles or by active, beating motion of projections from the surface of a living organism\cite{Marchetti:2013pi}.  The paradigm of such swimming motion is electrophoresis, driven by an external electric field on a charged body\cite{Delgado:2007qy}.  In ordinary fluids any such body is surrounded by ions that cancel its net charge, thus cancelling any net force due to the external field.  Still, the opposing forces on the surface and the nearby screening ions create a relative motion between the surface and the fluid.  The body moves forward by pushing the fluid backward.  

Individual swimming bodies such as electrophoretic colloids can show complex and controllable motion. A body can assume chiral steady-state rotation which can be synchronized with other like bodies by suitable external driving\cite{AjdariLong,Eaton-Moths-Witten,Moths-Witten1,Moths-Witten2}.  This is in addition to the striking forms of cooperative motion---such as swarming---arising from interparticle interactions\cite{Marchetti:2013pi}.    Such motions are of increasing interest as reproducible, asymmetric colloidal bodies become increasingly available\cite{Sacanna:2011fk,Meng560}.  Here we demonstrate a new method of calculating these motions for swimming bodies driven by electrophoresis.  The method is applicable to a broad range of colloidal swimming mechanisms.  

Though  nonlinear electrophoretic responses have recently been developed with dramatic effects\cite{Bricard:2013pr}, we focus here on the simplest linear response to the field.  Further, we consider the simple regime of strong screening. 
Strong screening means that the electrostatic screening length is arbitrarily small on the scale of the curvature of the body.    
The effects we aim to capture are from the body's shape and from its charge distribution.  

Our method exploits J. L. Anderson's insightful representation of the electrostatic flow over a surface\cite{AndersonSpheres}.  At any point of the surface that bears charge, there is a nonzero slip velocity proportional to the external field.  The flow velocity over a given point of the surface is solely determined by electrostatic forces near that point.  At such a point there is a local transverse surface electric field $\vector E^s$ proportional to the external field as perturbed by the non-conducting body.  This $\vector E^s$ depends on the shape of the body but not on the charge it bears.  To determine the flow velocity one needs only this surface field at the point in question times the ``zeta potential" between the bulk fluid and the charged body beneath\cite{AndersonSpheres}.  The electrophoretic motion of the body is then generated by this given velocity field as the sheath of fluid slips over the body.  
 
Though determining the slip velocity field is straightforward, inferring the resulting body motion is not. To determine this motion from the velocity at the surface is a challenging boundary value problem.   Below we describe a point source or boundary element method\cite{Read:1997fk, Youngren:1975rm} to determine this motion\footnote{
Boundary element methods for hydrodynamic flows aim to represent a smooth solid object by applying its boundary conditions at an array of points.  Our point source method only aims to represent a stokeslet object, \ie a rigid set of discrete drag forces.  It is more properly understood in the spirit of Kirkwood and Riseman \cite{Kirkwood-Riseman} with the additional imposition of rigidity\cite{Chen:1987jo}.  
}.  
We generate the needed velocity field using a set of $N$ point forces called stokeslets distributed over the surface. Each stokeslet produces a flow proportional to its force as dictated by the Oseen tensor Eq. \eqref{eq:Oseen}.  These stokeslets are sufficient the specify the surface velocity at $N$ points on the surface by solving a set of simultaneous equations.  The stokeslets create a flow outside the body consistent with the specified surface velocity relation\cite{Happel-Brenner}.    

The use of the Oseen tensor here implies the assumption that the body is at rest with respect to the distant fluid.  Holding the body at rest requires a net force and torque, which are transmitted to the fluid.  This net force and this torque are necessarily the sums of the stokeslet forces and torques that give the required surface flow, as determined above.  These are the constraint forces required to hold the body at rest.  

Knowing these stokeslet forces is sufficient to determine the motion when the body is released from rest.  Its velocity is simply the Stokes sedimentation velocity corresponding to the given force and torque.  Its angular velocity is given by an analogous rotation sedimentation mobility.  Imposing this velocity and angular velocity on the body necessarily generates a drag force and torque which cancels the electrostatic force and torque calculated above.  The result is that no force or torque is transmitted to infinity, as required for electrophoresis.  This simple superposition of sedimentation drag and electric effects is possible because the electrostatic slip velocity relative to the body is not affected by overall motion of the body, as recognized by Anderson\cite{AndersonSpheres}.  

In the next section, Section \ref{sec:implementation}, we spell out our implementation of this scheme.  Section \ref{sec:validation} describes our numerical tests for spheres, cubes and spherocylinders, confirming known results. In Section \ref{sec:chiral} we discuss how chiral motion arises in terms of the two tensors that give the velocity and angular velocity.  In Section \ref{sec:ChiralLocalized} we give quantitative predictions of chiral motion for specific shapes.  Even shapes as symmetric as a cube are shown to give substantial chiral response.  In the Discussion Section ( \ref{sec:discussion}) we discuss experimental implementations, concluding that these effects are readily observable despite potential limitations.  We discuss how the distinctive responses of asymmetric bodies can be used, noting how chiral response allows novel ways to manipulate the orientations of bodies via time-dependent applied fields.  Finally, we discuss how our method may be generalized to other forms of driving.    
 
\section{Point-source implementation}\label{sec:implementation}

\begin{figure}[htbp]
\includegraphics[width=.6\columnwidth]{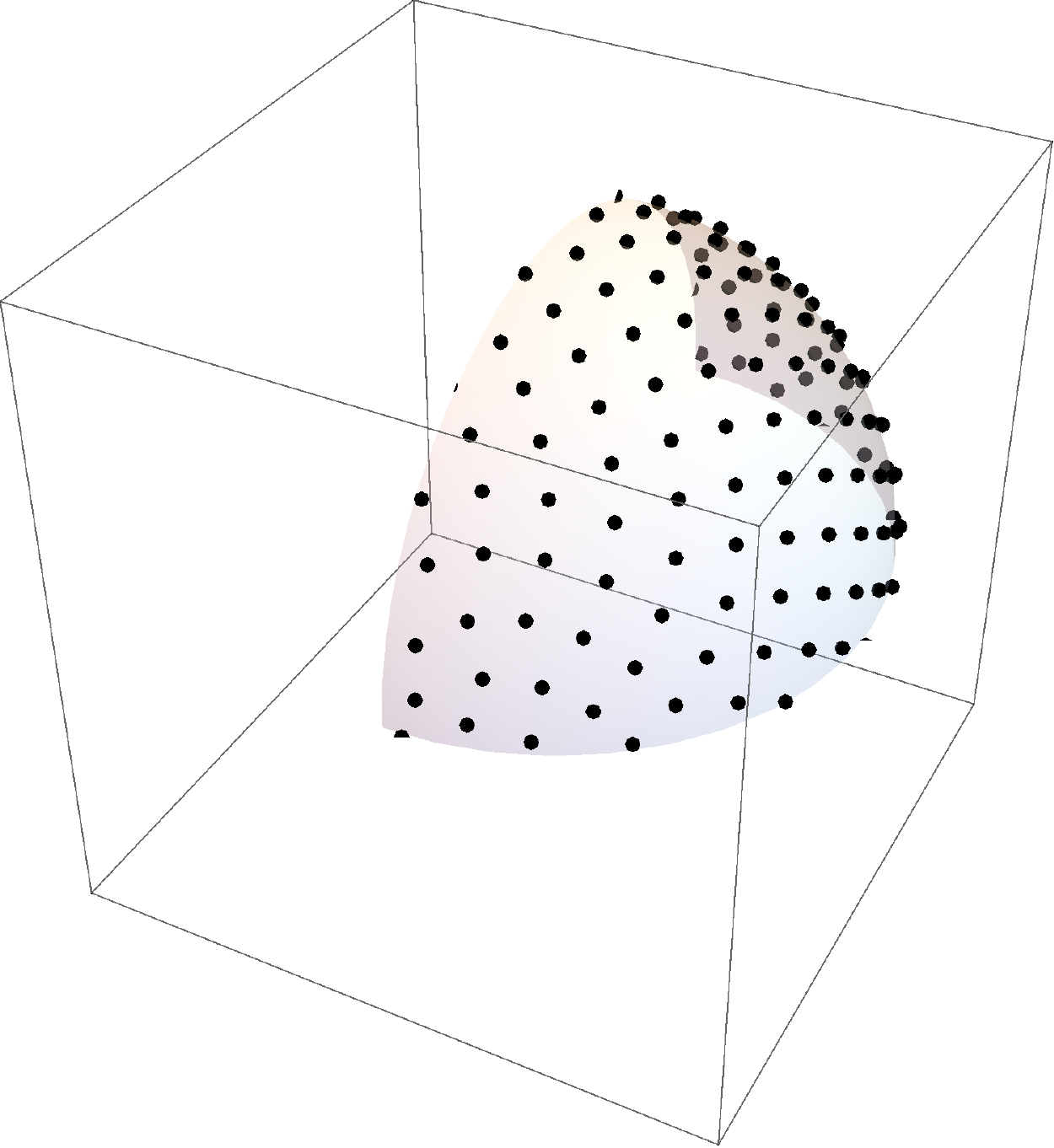}
\caption{\label{fig:dot_object}Representation of a solid sphere as a distribution of point sources used in Section \ref{sec:validation}.  
One quarter of the sphere is shown; the full sphere has 499 points.  We place the polarization charges $Q_i$ and stokeslet forces $\vector f_i$ at these points to generate the surface electric field $\vector E^s_i$ and velocity field $\vector v^s_i$. To determine stokes drag, a different set of stokeslet forces are determined at these same points.}
\end{figure}

Our use of superpositions of point sources is similar to our former work \cite{Mowitz:2017kx}.  We begin by defining a set  of mesh points labeled $i$ at which our various fields are to be sampled.  Several hundred mesh points at positions $\vector r_i$ are spread evenly over the surface as shown in Fig. \ref{fig:dot_object}.  For what follows, it is also necessary to know the normal unit vector $\hat n_i$ and the Voronoi area\cite{Okabe:2009bc} $A_i$ associated with each mesh point.

We then place a  small stokes sphere or stokeslet at each source point. An imposed set of forces $\vector f_i$ on these stokeslets generates a velocity field everywhere.  By constraining the stokeslets to maintain fixed relative positions, we may determine its rigid-body motion in an external field by the method of Kirkwood and Riseman\cite{Kirkwood-Riseman, Chen:1987jo}, detailed in the Appendix. We denote the set of fixed points at fixed mutual separations as a ``stokeslet object."  Analogously, we may place charges at the source points to create an electric field around the object.  We may choose these charges to implement a desired boundary condition on the electric fields at each stokeslet point under a given external field.  

We may consider this stokeslet object as a physical object which has a well-defined response to external forces or electric fields.  Once the stokeslet object is defined, these responses are uniquely determined by finite matrix operations to arbitrary accuracy, as discussed in the Appendix.  By choosing the stokeslet points to mimic the shape and charge distribution of a desired solid object, the stokeslet object's responses can also mimic those of the solid object to good accuracy, as shown below.  

As noted in the Introduction, we simplify the description by holding the charged object at rest and calculating the force exerted on the fluid as a result of the external electric field. 

      The numerical tasks needed are i) determination of the surface electric field, denoted $\vector E^s_i$ induced by a given imposed field $\vector E^0$, ii) determination of the slip velocity field $\vector v^s_i$, iii) determination of a set of stokeslet forces $\vector f_i$ that reproduce these $\vector v^s_i$, iv) determining the total force $\vector F$ and torque $\vector \tau$ resulting from these stokeslet forces, and v) finding the four Stokes mobility tensors that give the linear velocity $\vector U$ and angular velocity $\vector \Omega$ of the body for a given $\vector F$ and $\vector \tau$. 

\subsection{charge-independent aspects}\label{sec:chargeIndependent}
Of these quantities, the surface electric field i) and the Stokes mobility tensors v) depend only on the body's shape, not its charge distribution in the absence of the imposed $\vector E^0$.  Other numerical methods are available to do these tasks.  Our method enables us to describe the object in a common Stokeslet representation throughout the calculation.  For these tasks we follow the methods of Ref. \cite{Mowitz:2017kx} with little modification.  For definiteness we give the explicit equations in the Appendix.

In the Appendix we also summarize our calculation of the Stokes mobility tensors for the object as in Ref. \cite{Mowitz:2017kx}, using the method of Kirkwood and Riseman\cite{Kirkwood-Riseman, Chen:1987jo}.  This method can represent a hydrodynamically opaque object in which the interior fluid moves along with the body.  This calculation produces four tensors $\Mop_{V F}$, $\Mop_{V\tau}$, $\Mop_{\Omega F}$ and $\Mop_{\Omega \tau}$ such that
\begin{eqnarray}\label{eq:sedeqs}
\vector V =& ~\Mop_{V F} \cdot \vector F + \Mop_{V\tau }\cdot  \vector \tau \nonumber\\
{\rm and} &\nonumber\\
\vector \Omega =& ~\Mop_{\Omega F}\cdot \vector F + \Mop_{\Omega\tau} \cdot \vector \tau 
\end{eqnarray}

\subsection{charge-dependent aspects}\label{sec:chargeDependent}
The remaining tasks depend on the charge distribution or zeta potential over the object.  First we consider task ii): determing the surface velocities $\vector v^s_i$ at the mesh points. Each of these is determined by the Smoluchowski formula \cite{Smoluchowski:1903fv} using the local electric field $\vector E^s_i$ and zeta potentials $\zeta_i$: 
\begin{equation}\label{eq:vsi}
\vector v^s_i = -\zeta_i {\epsilon_r \epsilon_0\over \eta} ~\vector E^s_i  ,
\end{equation}
where $\eta$ is the viscosity of the fluid $\epsilon_r \epsilon_0$ is the dielectric constant, and $\zeta_i$ is the potential of the charged surface relative to the bulk (resting) solvent,
proportional to the surface charge density and the screening length.  
(A positively charged body with positive $\zeta$ moves towards the electric field, and the flow over the surface relative  to the body is away from the field.)  Since $\vector E^s_i$ was computed above and the $\zeta_i$ are presumed known, this formula determines the $\vector v^s_i$ and completes task ii).  

To address task iii) we determine the stokes velocity $v^s_{ij}$ at mesh point $i$ owing to a stokeslet at mesh point $j \notequal i$ exerting a force $\vector f_j$ on the fluid.  This velocity is given by the Oseen formula. 
\begin{equation}\label{eq:Oseen}
\vector v^s_{ij} = {1 \over 8 \pi \eta}{\vector f_j + (\vector f_j \cdot \hat r) ~ \hat r \over |r|},
\end{equation}
where $\vector r \definedas \vector r_i - \vector r_j$ and $\eta$ is the viscosity. The imposed total $\vector v^s_i$ at stokeslet $i$ is then the sum of these $\vector v^s_{ij}$ over $j$, as detailed in the Appendix.  The resulting $3N$ equations give linear conditions sufficient to determine the $3N$ $\vector f$'s.  

Once the $\vector f_i$ have been determined, the total force $\vector F$ transmitted to the fluid is simply $\sum_i \vector f_i$.  Likewise, the total torque $\vector \tau$ about a given origin is $\sum_i \vector r_i \cross \vector f_i$.  As seen above, this force and torque are proportional to the external field $\vector E^0$.  By calculating these for a basis set of $\vector E^0$ we thus determine the matrices $\Mop_{FE}$ and $\Mop_{\tau E}$ defined by 
\begin{equation}\label{eq:FtauofE}
\vector F = \Mop_{F E} \cdot \vector E^0 ~;\quad 
\vector \tau = \Mop_{\tau E} \cdot \vector E^0
\end{equation} 
Determining $\Mop_{FE}$ and $\Mop_{\tau E}$  accomplishes task iv).  

At this point we have determined the force and torque applied to the object and transmitted to the fluid when the object is held at rest.  It remains to find the velocity and angular velocity of the object when released from rest.  This motion of the released body does not alter the electrophoretic force and torque calculated above; these are determined by the viscous drag across the slip layer, and they depend only on the relative velocity between the local surface and the adjacent screening charge\footnote{
For general stokeslet objects, this locality may not be well defined, since an arbitrary set of stokeslets need not resemble any smooth surface.  However, if stokeslets are arranged over a smooth surface with spacing much smaller than the local inverse curvature, the stokeslet object may approximate the corresponding smooth body, as noted above.  Then the above reasoning applies, a stokes mobility tensor may be determined, and the $\vector V$ and $\vector \Omega$ may be calculated\protect{\cite{Mowitz:2017kx}}
}.
Without constraint forces, these electric forces are balanced by drag forces due to the motion.  These drag forces themselves are just those that appear on the right side of Eq. \eqref{eq:sedeqs}. 

Combining Eq. \eqref{eq:sedeqs} with Eq. \eqref{eq:FtauofE}, we obtain the desired $\vector V$ and $\vector \Omega$ for given $\vector E^0$.  These have the form \cite{AjdariLong}
\begin{equation}\label{eq:eqsofmotion}
\vector V = \Mop_{VE} \cdot \vector E^0 ~;\quad
\vector \Omega = \Mop_{\Omega E} \cdot \vector E^0
\end{equation}
where
\begin{eqnarray}
\Mop_{VE} =&~ \Mop_{V F}\cdot \Mop_{F E} + \Mop_{V \tau}\cdot \Mop_{\tau E} \nonumber \\
\Mop_{\Omega E} =&~ \Mop_{\Omega F} \cdot \Mop_{F E} + \Mop_{\Omega \tau}\cdot \Mop_{\tau E}
\end{eqnarray}

This procedure generates motion of the stokeslet object with no addition of force or torque to the fluid, by construction.  It also obeys a discrete form\cite{Witten:2019fv} of the Lorentz Reciprocal relation\cite{Happel-Brenner}, adapted by Teubner\cite{Teubner:1982kq}.  Further, it may be used to represent solid objects to good accuracy, as we now show.

\section{Validation}\label{sec:validation}

To verify that our discrete source method is reliable in practice, we simulated several objects where we could validate the method against independent calculations.  We did extensive comparisons using a spherical object.  We also simulated a cube and a capsule shape to verify their behavior with uniform charge.  

\subsection{\label{sec:e-fieldRes} Sphere}
For our comparisons we used the 499-point stokeslet object pictured in Fig. \ref{fig:dot_object} and Table \ref{nonuniformmobilitytable}.  We first checked the accuracy of task i) by comparing our discrete-source values of $E^s/E^0$  against the known analytic formula.  The induced dipole moment resulting from $E^s$ differed by 3.8\% relative to the exact result.
\begin{table}[htp]
\begin{ruledtabular}
\begin{tabular}{| c | c | c | c |}
    \shortstack{Charge \\ distribution} & Predicted & Measured & Error \\ \hline
     & & & \\
        {\shortstack{Uniform Sphere}} & {$V_{xx}$ =  1.00} & 1.01 & 1\% \\ \hline & & & \\
    \raisebox{-.25\height}{\shortstack{Capped Sphere \\ \includegraphics[width=.15\columnwidth]{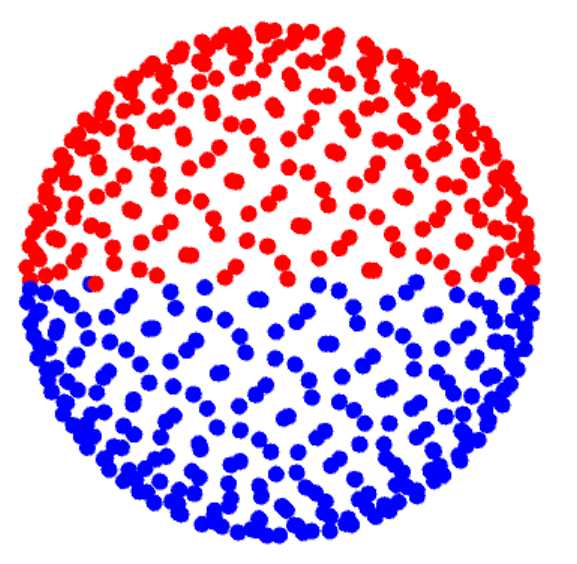}}} & {$\Omega_{yx}$ = 1.125} & 1.133& 1\% \\ \hline & & & \\
    \raisebox{-.5\height}{\shortstack{Striped Sphere \\ \includegraphics[width=.15\columnwidth]{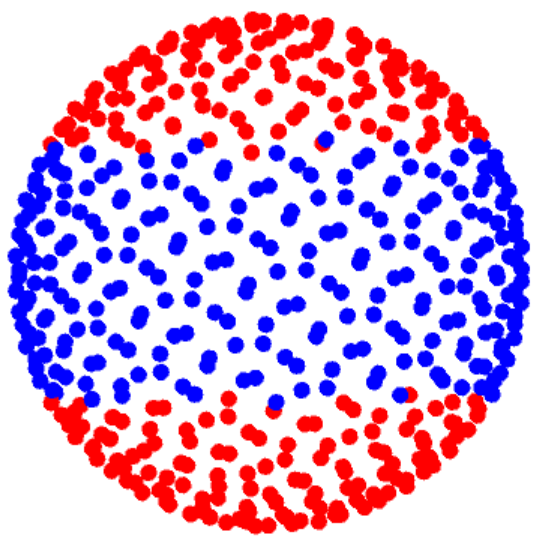}}}  & \raisebox{-.25\height}{\shortstack{$V_{xx}$ = -0.19 \\ $V_{yy}$ = -0.19 \\ $V_{zz}$ = 0.38}} & \raisebox{-.25\height}{\shortstack{ -0.18 \\-0.18 \\ 0.39}} & \raisebox{-.25\height}{\shortstack{\\$2.8$\% \\ $2.8$\% \\ $1$\% }} \\ \hline & & &\\
    {\shortstack{Uniform Cube \\ ~\\~ }} &  {\shortstack{$V_{xx}$ = $1.00$ \\ $V_{yy}$ = $1.00$\\$V_{zz}$ = $1.00$}} & {\shortstack{ 0.90 \\0.93 \\0.93}} &{\shortstack{$10$\% \\$7$\% \\ $7$\%}}\\ \hline & & &\\
  {\shortstack{Uniform Capsule \\ \includegraphics[width=.2\columnwidth]{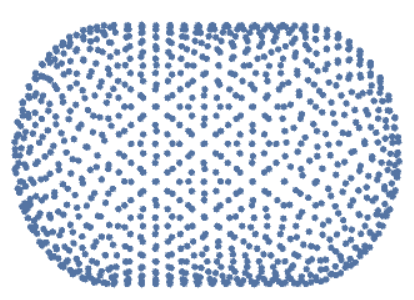}}} & {\shortstack{$V_{xx} = 1.00$ \\ $V_{yy} = 1.00$ \\$V_{zz} = 1.00$}} & 
   {\shortstack{ $1.05$ \\$1.06$ \\ $1.07$}} & {\shortstack{$5$\% \\ $6$\% \\ $7$\%}} 
    
\end{tabular}
\end{ruledtabular}
\caption{\label{nonuniformmobilitytable} Calculated electrophoretic motion for known cases.  The stokeslet object for the three sphere cases was the 499-point object pictured in Fig \ref{fig:dot_object}.  Stokeslets have a radius of 0.0252, 
and cover the sphere with an area fraction of 8 \%. The plane of the figure is the $x-z$ plane.  The notation $\Omega_{yx}$ indicates the $y$-directed angular speed (out of the page)  in a horizontal ($x$-directed) electric field, and similarly for velocities $V$.  Predicted velocities\cite{AndersonSpheres} are given in units of the Smoluchowski velocity of Eq. \eqref{eq:vsi} for the uniformly charged object.  Measured velocities are given in the same units.    The cube stokeslets are indicated in Fig. \ref{fig:CubeWithFs}.  The density in the y-z faces is slightly smaller than in the other faces; leading to a small anisotropy in the velocity response.  For each of the spheres the unreported forces and torques are consistent with 0 or are equal to the reported ones by symmetry.}
\end{table}

We then used our method to calculate the electrophoretic mobility of several charge distributions on a sphere.  Here we used the known analytic formula for $E^s$\cite{Griffiths:1999tg}.  The calculation was simpler and the results more accurate than our earlier version\cite{Mowitz:2017kx}.  We studied a uniformly charged sphere with zeta potential of $1$ and two nonuniform distributions.  One of these was a capped sphere where stokeslets in the top hemisphere had a zeta potential of $1$ and stokeslets on the bottom hemisphere had a zeta potential of $-1$.  The other was a striped sphere with stokeslets in the top and bottom quarters having a zeta potential of $1$ and the middle half having a zeta potential of $-1$, giving overall charge neutrality.
We observe that even using as little as 499 stokeslets, the motion due to an electric field in the three Cartesian directions are within 3\% of the expected electrophoretic mobility as shown in Table. \ref{nonuniformmobilitytable}. 

\begin{figure}[htbp]
\includegraphics[width=.48\columnwidth]{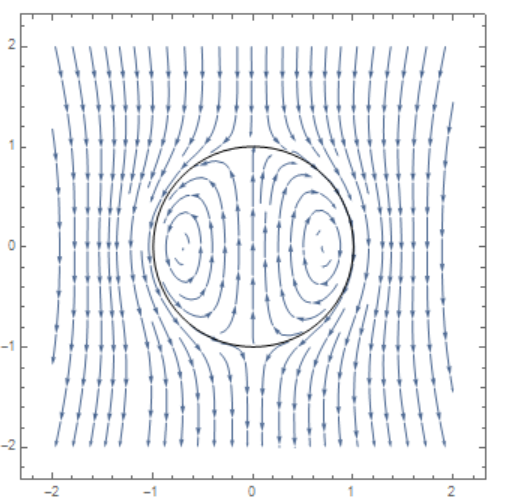}
\includegraphics[width=.48\columnwidth]{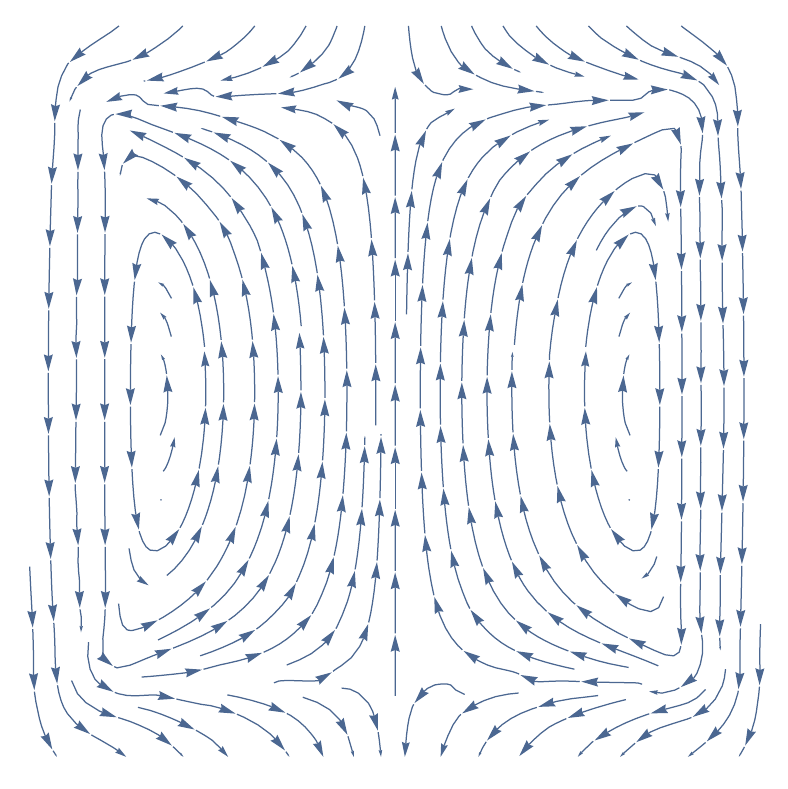}
\caption{\label{fig:flow} Flow fields calculated by the method of Sec. \ref{sec:implementation}.  Left: Flow around a uniformly positively charged, fixed sphere in a cross section across the equator.  The external field $\vector E^0$ is pointing up.  The black outline shows the position of the surface of the sphere. The flow lines show direction only, they do not show magnitude. Flow of the screening charge and hence the fluid is opposite to the field.  Right: analogous picture for a cube. }
\end{figure}

We also observed the total flow field created by the stokeslet forces $\vector f_i$'s. As expected, we see a tangential flow around the surface of the body. Additionally, there is a flow inside the body, since we do not use a solid body constraint in our calculations \footnote{
The interior flow is that which would occur if the slip velocity were imposed on a liquid sphere or cube.  It is not an artifact of the stokeslet discreteness.
}. 
The field for the uniform sphere is visible in Fig. \ref{fig:flow}. While not apparent from the field diagram, the flow in the figure falls off inversely with distance as expected from stokes flow.

\subsection{\label{sec:boxValidation}Cube}
We calculated the electrophoretic motion of a cube represented by 1542 stokeslets, as shown in Fig. \ref{fig:CubeWithFs}.  We represented each face as a regular lattice of points.  Maintaining continuity at the edges while maintaining a symmetric cubical shape required slight anisotropy of the lattice on the different faces.  While flat-sided shapes simplify the uniform placement of points on each face, they complicate the treatment of edges. Our calculation requires assigning a normal direction to each point.  Thus we omitted the edge points in our cube, which have no well-defined normal.  Our calculation also requires an assigned area for each point.  These areas varied in our cube, especially next to the edges.  To obtain well-defined areas we numerically determined the Voronoi area for each point.  

\begin{table}[h]
\begin{tabular}{| c | c | c | c |}
    \hline
    $\vector E^0$ & Dipole moment & Predicted Dipole Moment &  Error \\ \hline \hline
    \{0,0,1\} & \{0,0,-0.0804\} & \{0,0,-0.0795\} & 1.2\% \\ \hline
    \{0,1,0\} & \{0,-0.0804,0\} & \{0,-0.0795,0\} & 1.2\% \\\hline
    \{1,0,0\} & \{-0.0794,0,0\} & \{-0.0795,0,0\} & 0.1\%\\ \hline
\end{tabular}
\caption{\label{dipoleMoments} The depolarization field created to counteract the normal component of $\vector E^0$ creates a dipole in the direction of $\vector E^0$. Here the dipole moments for a cube with $\vector E^0$ in each of the cartesian directions are recorded. The calculated dipoles are identical in two directions, however due to the varying density of depolarization charges on two of the six faces of the cube, the dipole moment in the third direction is slightly ($~1$ \%) different.  The predicted dipole moments are 11\% greater than the dipole moments of the depolarization field of a sphere with the same volume. This correlation was calculated in \cite{Herrick:1977px}. 
}
\end{table}

By assigning a unit zeta potential to the stokeslets, we could compare the calculated electrophoretic speed with the Smoluchowski prediction.  The calculated speed was 8-10\% too small depending on orientation.  The observed anisotropy arises from the different arrangement of the points on different faces of the cube.  To understand the overall discrepancy, we checked the Stokes sedimentation mobility \cite{Heiss:1952zl} and the induced electric dipole moment of the cube(Table \ref{dipoleMoments}) against published calculations. Both showed only small discrepancies from the predications.  We further verified that the flow velocity around the cube falls off as the inverse third power of the distance at large distances, as expected for  electrophoretic motion.  We observed the flow generated by the stokeslet forces $\vector f_i$ (Fig. \ref{fig:flow}.  As with the sphere we saw the required tangential flow near the boundary of the body  and a flow inside the body. 

A remaining aspect that we could not check was whether the imposed surface velocities $\vector v^s$ at the stokeslets sufficed to represent the expected potential flow\cite{Morrison} around a uniformly-charged object.  Since our velocities necessarily change abruptly at the edges of our object, it is plausible that our discrete representation is deficient in this respect.  Our improved agreement with the smooth capsule shape reported below supports this view.  



\subsection{\label{sec:pillValidation} Capsule shape}
To evaluate the accuracy of representing smooth objects of lower symmetry, we studied the 1542-stokeslet 
spherocylinder or capsule shape of Fig. \ref{fig:PillTwoChargeSymmetric}a. As with the cube, we assigned a unit zeta potential to each Stokeslet. The translational velocity thus obtained was only 5-7\% different from the expected Smoluchowski velocity. This velocity varied by no more than $2 \%$ in different orientations despite substantial anisotropy of the capsule.  Unlike for the cube where the velocity was smaller than expected, the capsule travelled faster than expected.  We attributed these higher speeds to inaccuracy in determining the Stokes sedimentation mobility.  This calculation requires good exclusion of the external flow from the interior of the object, but we noticed incomplete exclusion where the density of points was low, so that the external flow is like that of a smaller object.   We verified that the velocity field around the object varies smoothly away from the object, as with the sphere and the cube.

\section{Chiral responses}\label{sec:chiral}
					The linear relations of Eq. \eqref{eq:eqsofmotion} determine the motion, \ie the time dependence of $\vector V$ and $\vector \Omega$.  This motion can be subtle since the $\Mop$'s depend on orientation and are thus influenced by the calculated $\vector \Omega$.  To work out this dependence, it suffices to consider the $\vector \Omega$ equation. Once $\vector \Omega(t)$ is found from this equation, the matrices $\Mop_{VE}$ and $\Mopo$ are known functions of time and $\vector V(t)$ may be inferred immediately. 
					
As the body rotates with angular velocity $\vector \Omega$, any matrix $\Mop$ characterizing it rotates together with the body.  Denoting $\Rop(t)$ as the rotation matrix from the lab frame to the body frame at time $t$, the matrix $\Mop(t+dt)$ in the lab frame is then\cite{Marion:2013kx}:
\begin{equation}\label{eq:MopRop}
\Mop(t+dt) = \Rop(dt) \cdot \Mop(t)\cdot  \Rop^T(dt) 
\end{equation}
Here $\Rop(dt)$ is a differential rotation related to $\vector \Omega$ by the antisymmetric matrix denoted $\xop{\Omega}$ defined by
\begin{equation}
\xop \Omega \cdot \vector A \definedas \vector \Omega \cross \vector A
\end{equation} for any vector  $\vector A$.  
Specifically, 
\begin{equation}
\xop{\Omega} \definedas ~ - ~\left(
\begin{array}{ccc}
 0 & \Omega_3  & -\Omega_2  \\
 -\Omega_3 & 0  & \Omega_1  \\
 \Omega_2 & -\Omega_1  & 0  
\end{array}
\right) .
\end{equation} 
Now $\Rop(dt)$ and $\Rop^T(dt)$ can be written
\begin{eqnarray}
\Rop(dt) =& ~\IdentityMatrix + dt~ \xop{\Omega}~ ; \nonumber\\
\Rop^T(dt) =& \Rop(-dt) .
\end{eqnarray}
Using these relations, we infer
\begin{equation} 
\dot\Mop(t) =  \xop{\Omega}\cdot \Mop(t) ~ - ~ \Mop(t)\cdot \xop{\Omega} 
\end{equation}
or, in commutator notation
\begin{equation}\label{eq:Mopdot} 
\dot\Mop = [\xop{\Omega}, \Mop(t)] 
\end{equation}
Recalling that $\vector \Omega = \Mopo\cdot \vector E^0$, the equation of motion for $\Mopo$ is evidently
\begin{equation}\label{eq:Mopodot}
{d\over dt} \Mopo = [(\Mopo\cdot \vector E^0)^\cross, \Mopo ]
\end{equation}

					Rotational motion of the form of Eq. \eqref{eq:Mopodot} is encountered in several contexts.  One is sedimentation of an asymmetric body under an external force $\vector F$. Here $\Mopo$ is replaced by $\Mop_{\Omega F}$ of Eq. \eqref{eq:sedeqs} and $\vector E^0$ is replaced by $\vector F$.  Another is the free rotation of a rigid body with conserved angular momentum $\vector L$ and inertia tensor $\Iop$\cite{Marion:2013kx}.  Here the equation of motion has the form $\vector \Omega = \Iop^{-1}\cdot \vector L$, so that $\Mopo$ is replaced by $\Iop^{-1}$ and $\vector E^0$ is replaced by $\vector L$.  Here the tensor of interest is symmetric.  We begin our discussion of the electrophoretic motion by considering the analogous case of a symmetric $\Mop_{\Omega E}$ tensor

\begin{figure}[htbp]
\includegraphics[width=\columnwidth]{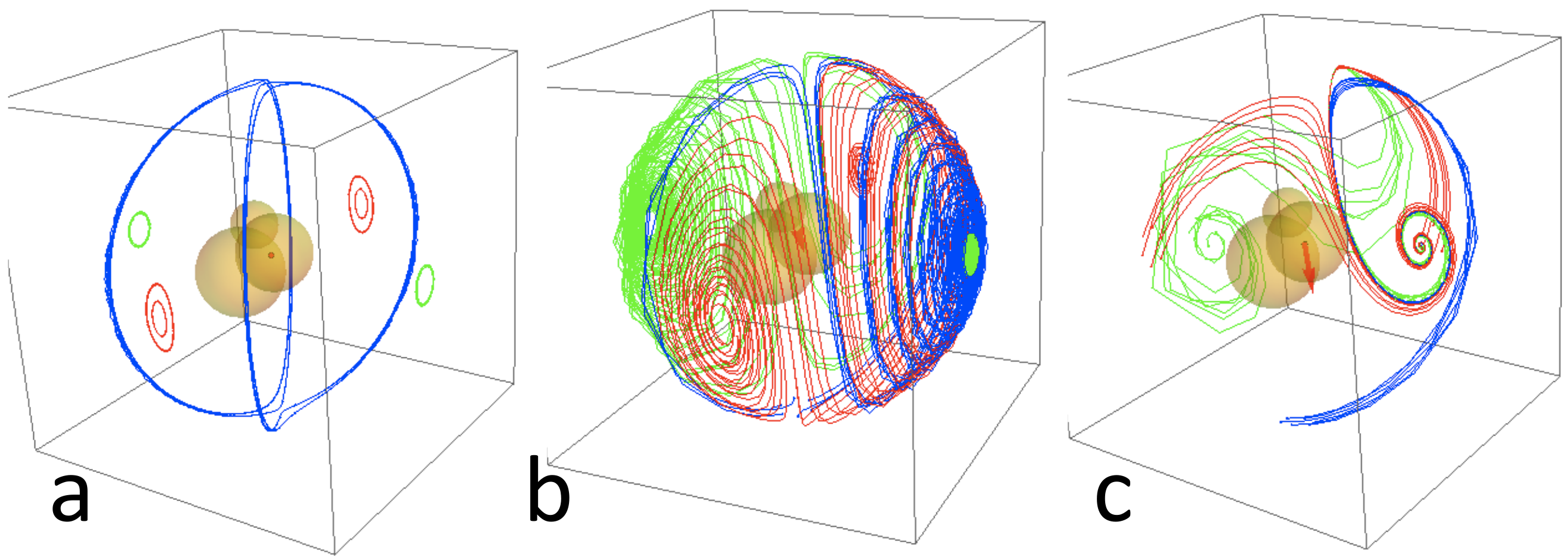}
\caption{\label{fig:Orbits} 
Example motions induced by $\Mopo$ showing the effect of the antisymmetric part. Colored lines are trajectories traced by $\vector E^0$ as viewed in the frame of the object, beginning near each fixed point. Color of trajectory indicates which eigen axis it started from.  a) no antisymmetric part.  All trajectories are closed. Fixed points are at the center of each cube face.  Two pairs of trajectories (red and green) remain localized near their starting points. Trajectories starting near the bottom and top fixed points (blue) form a single connected trajectory that oscillates between the two starting points.  b) small antisymmetric part.  Trajectories starting near the three unstable fixed points (front, left and bottom) spiral away from their starting point.  All converge to the stable fixed point on the right.  Trajectories near the right and rear stable fixed points converge to the local stable fixed point. Since the motion of any point on a trajectory depends only on its location on the sphere, no two trajectories may cross.  c) large antisymmetric part.  All starting points converge to the stable fixed point on the right. Trajectory from top unstable fixed point not shown.}
\end{figure}

					A symmetric $\Mop$ has three orthogonal eigen axes and hence six eigen directions.  When $\vector E^0$ is aligned with one of these, it remains so aligned and thus $\vector \Omega$ remains fixed.  When $\vector E^0$ is slightly displaced from one of these directions, it doesn't systematically return to it. That is, the fixed $\vector\Omega$'s are at most neutrally stable.  Figure \ref{fig:Orbits}a shows an example.  

					If one perturbs such a symmetric $\Mop$ with a small antisymmetric addition,  the equivalence of the positive and negative eigendirections is broken.  One of this pair of fixed points becomes locally stable while the other is unstable\cite{Moths-Witten2}. The eigendirections $\hat n$ in which $\Mopo ~\hat n = \lambda_n \hat n$ also shift and are no longer orthogonal.  Now typically $\vector \Omega$ evolves to one of the locally stable fixed points.  Thus the long-time motion is rotation around a stable eigendirection such that  $\vector \Omega~  (= \Mopo\vector E^0)$ and $\vector E^0$ are parallel.  The final angular velocity $\vector \Omega_f$ is given by the corresponding $ \lambda_n \vector E^0$.  

					The opposite extreme is a purely antisymmetric $\Mop$.  Any antisymmetric $\Mop$ can be written in the form $\Mop = (~\vector p~)^\cross$ for some vector $\vector p$ denoted the ``dipole vector".  The unit vector $\hat p$ rotates in time according to 
\begin{equation}
\dot{{\hat p}} = \vector \Omega \cross \hat p = (\xop{p} \cdot \vector E^0) \cross \hat p = (\vector p \cross \vector E^0) \cross \hat p
\end{equation} 	
This simplifies to 
\begin{equation}
\dot{{\hat p}} =  p E^0 \left(~\hat E^0 - \hat p ~(\hat p \cdot \hat E^0 )~\right) .
\end{equation}
This is just the equation for an electric dipole relaxing in the external field $\vector E^0$. The quantity $\hat p \cdot \hat E^0$ is strictly increasing with time except when $\hat p \parallel \hat E^0$; thus any initial $\vector p$ reaches a final state aligned with $\vector E^0$.  
Evidently $\hat p$ is the only real eigen axis of $\Mopo$, and its eigenvalue is 0. 

					 The same behavior occurs if a small symmetric part is added to this $\Mop$; there is only one real eigenvalue $\lambda_1$. This $\lambda_1$ is no longer zero in general.  One direction on this axis is globally stable; any initial $\vector \Omega$ evolves to this eigendirection\cite{Gonzalez04}. A body with this property evidently has a preferred direction of rotation around $\vector E^0$.  It thus shows a clear chirality.  We call such bodies axially aligning.  The property of axial alignment offers a kind of handle allowing a set of such bodies to be manipulated\cite{Moths-Witten1,Moths-Witten2}.  

					 For almost all $\Mopo$ matrices one may achieve this axially aligning behavior by multiplying the antisymmetric part by a sufficiently large factor\cite{Krapf:2009ai}. This suggests that among asymmetrically charged bodies, axial alignment is not uncommon.  Whether axial alignment of charged bodies is appreciable in practice is an open question.  How strong can axial alignment be?  What conditions are necessary to create it?  

					Axial alignment requires conditions on both the shape and the charge distribution of the body.  As for charge distribution, the Morrison theorem\cite{Morrison} guarantees that a {\em uniform} distribution gives no rotation: $\Mop_{\Omega E} = 0$ regardless of its shape.  As for shape, it is known\cite{AndersonSpheres} that a spherical shape cannot be axially aligning with a nonzero rotation frequency, regardless of its charge distribution.  However, preferred chirality does not require a {\em chiral} shape, as shown by Ajdari and Long\cite{AjdariLong}. It is sufficient for the charge distribution to be chiral.  Though the shape need not be chiral, it is not known what departure from a spherical shape is needed.  

\section{Chiral motion from localized charges}\label{sec:ChiralLocalized}
					
					Using the methods of Sections \ref{sec:implementation} and \ref{sec:validation}, we may readily explore the range of chiral behavior obtained with simple shapes.  In this section we show that strong chirality can occur even with no special regard for the shape.  We consider the cube shape and the capsule shape of Sec. \ref{sec:validation}.  Evidently \footnote{
Each point source $\zeta_i$ gives rise to a surface flow field that is proportional to $\zeta_i$ at stokeslet $i$ and zero elsewhere.  Thus the overall surface flow field $\vector v^s_i$ is the superposition of the contributions from each $\zeta_i$. Finally, the force, torque and hence the velocity and angular velocity are linear in the surface flow field.  Thus the nonlinear properties of electophoresis result from nonlinear dependence on the position and drag coefficients of the stokeslets, not on their charges.  
}
$\Mopo$ is the sum of the response matrices from each point on the object, \ie the sum of contributions from each $\zeta_i$.  Accordingly, we consider objects with isolated points of charge.  First we recall the factors that limit chiral behavior and set its scale.

					 To estimate the magnitude of chiral rotation, it is natural to use the typical scales of velocity found in experiments.  These have electric fields of the order of 100 volts/cm and zeta potentials of the order of tens of millivolts.  This implies nominal electrophoretic speeds of order 100 microns/sec.  Thus a natural scale for an angular velocity in electrophoresis is such that the surface velocity is the Smoluchowski velocity of Eq. \eqref{eq:vsi} for the system in question.  In what follows we will compare angular velocities in this spirit, in terms of Smoluchowski speed divided by a characteristic body dimension.   
					 
					To estimate the effect of nonuniform charge on the overall scale of the motion, we may compare with the case of a sphere.  Here only the monopole and quadrupole moment of the zeta potential affect the translational motion, and only the dipole moment affects the angular velocity\cite{AndersonSpheres}.   For the simple case of bodies with a single sign of charge, these moments are of comparable order, barring special symmetries.  Thus to simplify comparisons, we may consider bodies with a single sign of charge and with the same total charge.  The $\Mop$'s for an object with both positive and negative charges are then simply the sum of a term for the positive charges and a second term for the negative charges.

\subsection{single point charge}\label{sec:pointcharge}
\begin{figure}[htbp]
\includegraphics[width=\columnwidth]{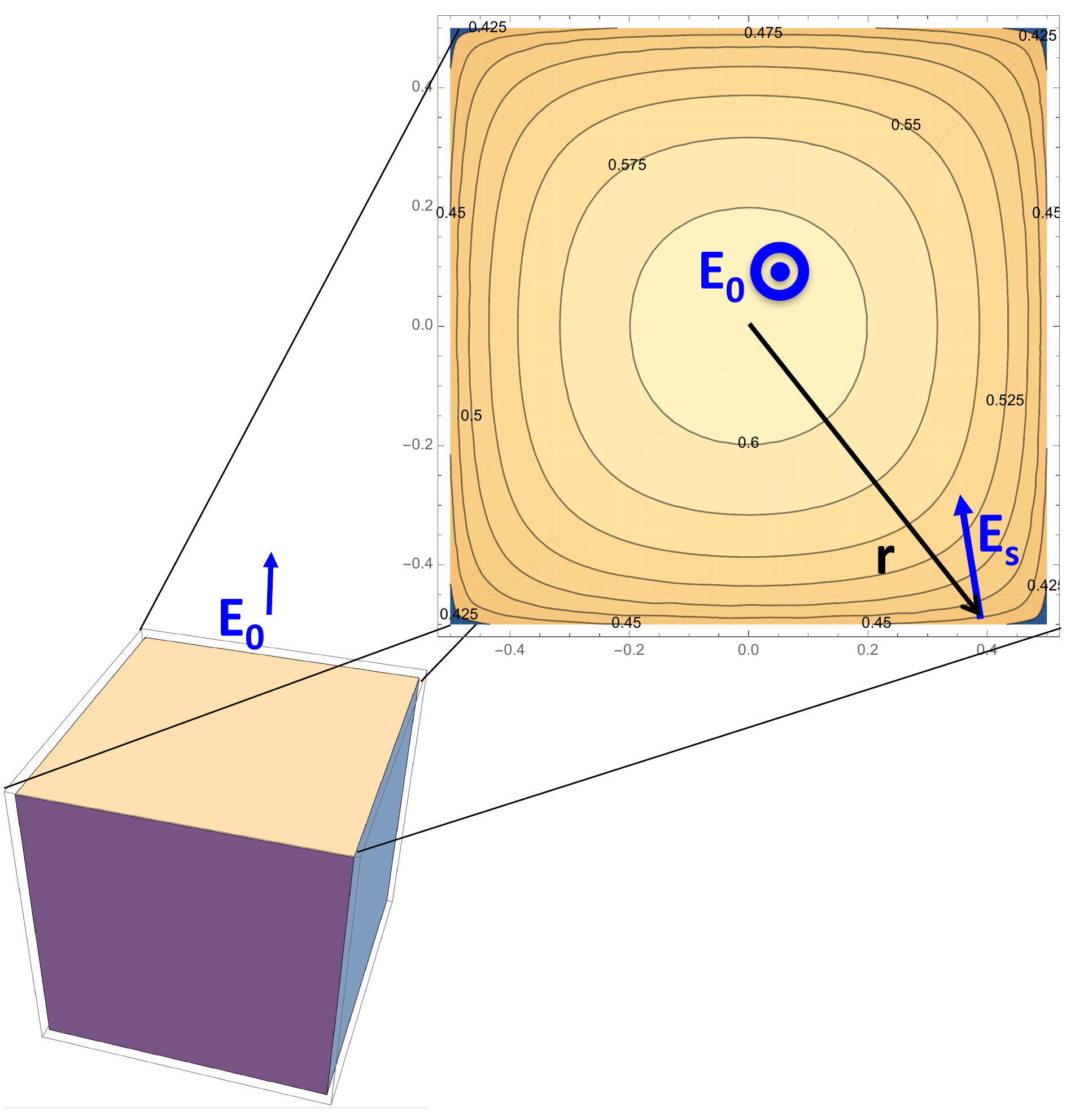}
\caption{\label{fig:CubeWithEs} 
Electric fields on a cube. Lower left: perspective view of cube with upward-pointing external field vector $\vector E^0$.  Upper right: top view showing contour lines of electrostatic potential using a commercial software package\cite{ANSYSMaxwell}, courtesy of Gerwin Koolstra. The $\vector r$ points from the center of the cube to a site near the right corner of the front edge.  $\vector E^s$ indicates the direction of the surface field there, perpendicular to the isopotential lines. This $\vector E^s$ has a component perpendicular to both $\vector r$ and $\vector E^0$.}
\end{figure}

We first consider objects where positive charge is confined to an arbitrarily small region of the surface, so that it may be treated as a point charge.  For this case the structure of $\Mopo$ is simplified.  The case of a point charge on a sphere shows the overall behavior. We suppose that $\vector E^0$ is upward and that a positive point charge is initially on the horizontal equator of the sphere.  The negative screening ions near the point charge are pushed vertically downward by the upward surface field $\vector E^s$, and the adjacent surface is pushed upward.  This push leads to a rotation of the charge vertically upward.  The force on the rotated charge continues to push it upward.  But when the charged has reached the top of the sphere, there is no tangential surface field to push the screening charge.  Accordingly, the motion stops.  

This behavior generalizes to arbitrary shapes and arbitrary charge locations.  As with the sphere treated above, $\Mopo$ is proportional to the surface field $\vector E^s$ at the charged site $i$.  This $\vector E^s$ is in turn proportional to $\vector E^0$ via a matrix $\Mop_{E^s~ E}$.  These $\vector E^s$ at $i$ are restricted: they must lie in the two-dimensional tangent plane at $i$ for any three-dimensional $\vector E^0$.  Thus $\Mop_{E^s~ E}$ cannot be invertible: it must have at least one null vector, denoted $\hat O$, for which $\Mop_{E^s~ E} \cdot \hat O$ = 0.  (This null direction need not be in the normal direction $\hat n_i$ as in the case of the sphere.)  Any $\vector E^0$ in this direction can give no rotation, since there is no surface field to drive motion.  Indeed, no translation can occur for this $\vector E^0$ either.  


Since $\vector E \parallel \hat O$ can give no rotation, it is necessarily a fixed point of the dynamics.  We have noted that whenever $\Mopo$ has a unique real eigenvalue, its eigendirection must be a globally stable fixed point \cite{Gonzalez04}.  The single-point-charge examples examined below $\Mopo$ indeed had a unique real eigenvalue and thus the fixed point was globally stable.  Thus for these examples the only field that can produce motion---$\vector E_s$---vanishes and all motion must come to a stop.  Thus a single charge does not generically give ongoing chiral motion.  In view of this finding we are led to consider objects with two point charges. 

\subsection{two point charges}\label{sec:twopointcharges}
In order to find a final state of steady chiral rotation we require an $\vector E^0$ such that $\vector \Omega$ is along $\vector E^0$.  We first consider an object with symmetrical shape whose drag tensors are isotropic, such as a cube.  We ask whether it is possible to place a pair of charges so that there is chiral rotation.  Since the body's shape is isotropic,  $\vector \Omega$ and the torque $\vector \tau$ must be parallel.  Thus the $\vector E^0$ must give a torque parallel to $\vector E^0$.  This torque is necessarily the sum of the torques due to the two charges.  Either of these torques may have components not parallel to $\vector E^0$, but these components must be equal and opposite.  

Each of these torques must come from a local force from each of the charges.  In order to create a torque along the $\vector E^0$ axis, there must be a force perpendicular to $\vector E^0$ and to the ``moment arm" $\vector r_i$  from the center of drag to the  charge, as shown in Fig. \ref{fig:CubeWithEs}.  The direction of this force is dictated by the direction of the surface field $\vector E^s$.  The needed torque would be consistent with a surface field component perpendicular to $\vector E^0$ and to $\vector r_i$.  As shown in the figure, a cube with $\vector E^0$ along one axis has points $\vector r_i$ with this property.  

As noted in Sec. \ref{sec:pointcharge}, a single charge at such a point need not give chiral rotation.  Instead, it may simply rotate into the fixed-point direction in which the surface field vanishes.   To avoid this outcome, there must be other torques so that the full total torque is along $\vector E^0$.  As seen in the Fig. \ref{fig:CubeWithEs}, two like charges placed at opposite points near the corners of a face fill the requirements.  

Here we have argued that the needed torques can arise from the surface fields of a cube.  To verify this requires a calculation following the methods of Sec. \ref{sec:implementation}.  We describe examples in the next subsection.

The above reasoning suggests a prescription for creating chiral rotators with pairs of like point charges.  One chooses an axis of symmetry as the desired direction of $\vector E^0$.  Then with $\vector E^0$ in this direction one identifies points $\vector r_i$ where symmetry allows the surface field to have a component perpendicular to both $\vector E^0$ and $\vector r_i$.  Finally one places a second point at an opposite position such that the two surface fields sum to a vector along $\vector E^0$.  In the next section we implement this prescription with a capsule shape.

\subsection{Examples}\label{sec:examples}
The above examples suggest ways to arrange charge that rotate about a particular axis in a given chiral sense.  The argument neglects many specifics, such as the precise relation between the surface field and the torque. Thus they give no quantitative measure of the effect. Neither do they address the behavior of the object in other orientations.  In this section we explore these questions using two nonchiral shapes: a cube and a capsule shape each with one or two point charges.  

\subsubsection{Cube}\label{sec:cube}
\begin{figure}[htbp]
\includegraphics[width=.75\columnwidth]{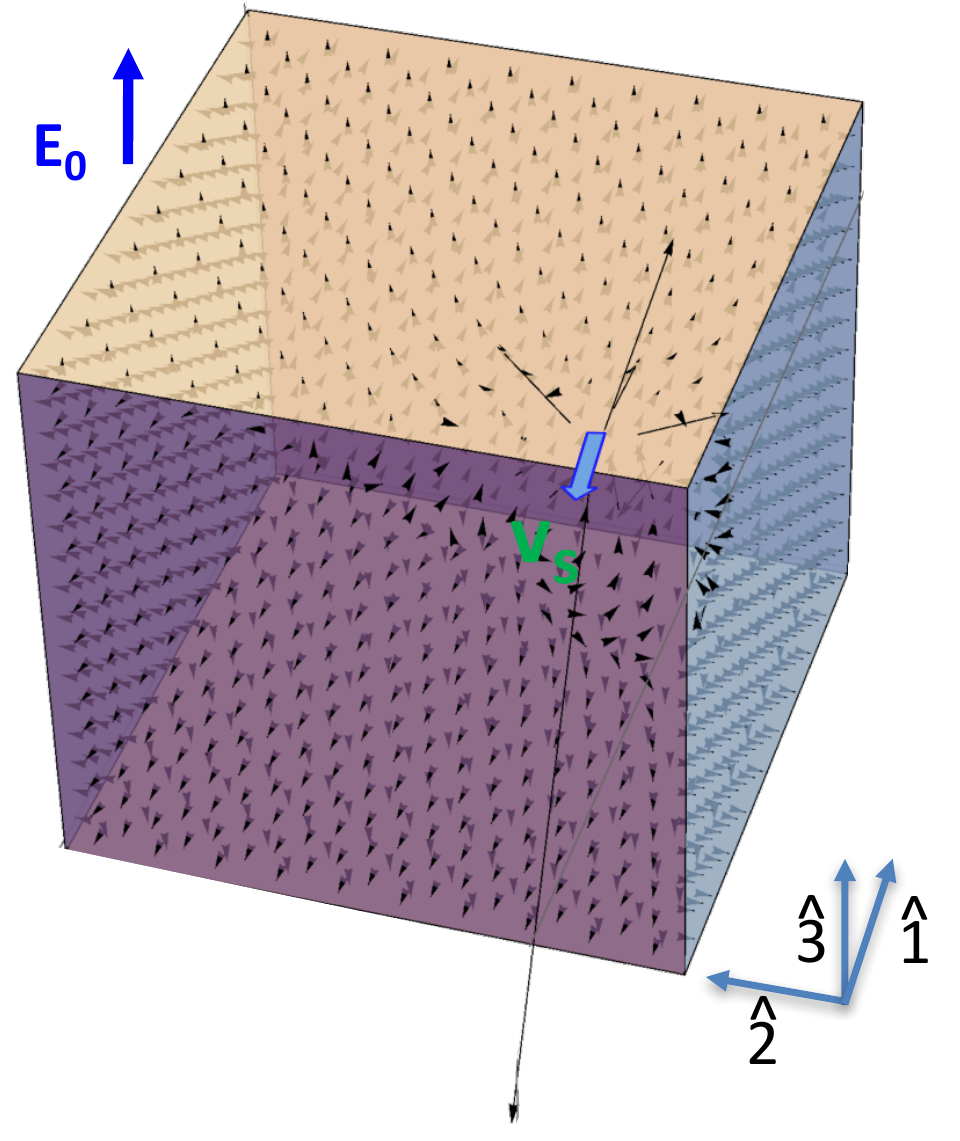}
\caption{\label{fig:CubeWithFs} 
Stokeslet forces on a singly charged cube. Upward external field $\vector E^0$ in $(1,0,0)$ direction is indicated. Colored arrow shows the direction of surface velocity $\vector v^s$ at the charge position (\cf Fig. \ref{fig:CubeWithEs}). Small black arrows show stokeslet forces needed to create the indicated $\vector v^s$ at the charge and 0 elsewhere, these forces are concentrated near the charge.  Basis at lower right shows the $\hat 1, \hat 2$ and $\hat 3$ axes used for the $\Mopo$ and $\Mop_{VE}$ matrices reported in the text.}
\end{figure}

We used the 1542-point stokeslet unit cube described in Section \ref{sec:validation}.  We first gave the cube a single charge at the position shown in Fig. \ref{fig:CubeWithFs} on the $(x, y, 1)$ face next to the $(0, y, 1)$ edge.  Implementing the needed $\vector v^s$ field produced the stokeslet forces indicated.  Enforcing the $\vector v^s{}_i = 0$ condition away from the charged site generated strong stokeslet forces opposing the one at the charged site.  The velocity field away from the stokeslet sites varied strongly from site to site, but became smooth beyond a fraction of the cube length.  

\begin{figure}[htbp]
\includegraphics[width=\columnwidth]{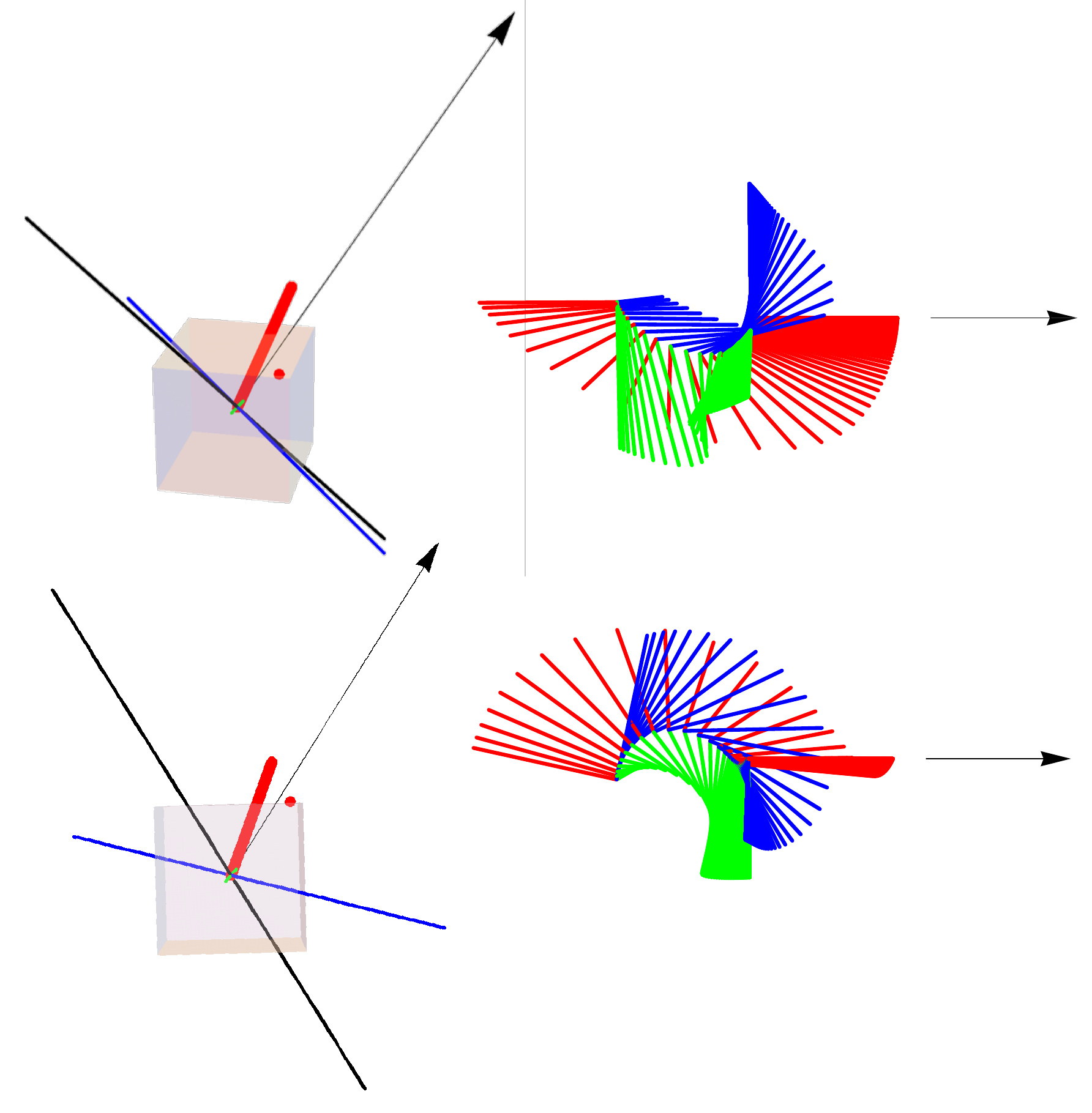}
\caption{\label{fig:1ChargeCubeAxesTrace} 
Left: Characteristic axes of the $\Mopo$ matrix for the unit cube with a single unit charge.  Position of charge is shown as a colored dot.  Top left:  view from the perspective of Fig. \protect \ref{fig:CubeWithEs}. Heavy colored bar has unit length and marks the aligning direction.  Arrow shows the dipole vector.  The two perpendicular axes are the principal axes of the symmetric part of $\Mopo$ with lengths equal to the eigenvalues in the units defined in the text.  The third much shorter axis is nearly parallel with the aligning direction. Bottom left: front view. Right: multiple-exposure views of cube motion in right-pointing external electric field $\vector E^0$ indicated by an arrow, calculated as described in the text.  The cube is indicated by an orthogonal basis with the light colored (red) axis in the aligning direction (\cf upper left drawing).  Length of the axes is the cube size. 100 exposures are shown.  The time between adjacent exposures is the time for the cube with unit charge spread uniformly to travel 0.05 cube lengths.  
Initial orientation was slightly displaced from the negative aligning axis, an unstable fixed point.  Motion accelerates away from this fixed point as the cube rotates towards the stable fixed point and translates to the right.  Exposures collapse on the right as translation and rotation slow to a stop as explained in Sec. \ref{sec:pointcharge}.  Lower picture shows the same motion from a view angle rotated 90 degrees about $\vector E^0$ relative the upper picture.
}
\end{figure}

Clearly the net force on the object is more complicated than the single point force considered in Secs. \ref{sec:pointcharge} and \ref{sec:twopointcharges}.  Still, the force is localized near this charge.  For a point shear force applied near a flat surface, the drag force falls off as the -3 power of distance\cite{Cortez:2015pt}.  Much of the force at the charge is thus cancelled by nearby forces, creating a strong force dipole.  Yet some of this force must survive for a finite object, since there is a nonzero electrophoretic velocity.  Thus the localized force picture of Secs. \ref{sec:pointcharge} and \ref{sec:twopointcharges}, is qualitatively consistent with observations.

Following the procedure of Sec. \ref{sec:implementation}, we computed the translation mobility $\Mop_{VE}$ relative to that of the corresponding uniformly charged cube is 
\begin{equation}\label{eq:MopVE1chargecube}
\Mop_{VE} =
\left(
\begin{array}{ccc}
 0.773804& -0.312964& -0.10127\\
 -0.242914& 1.94923& -1.14369\\
 0.00108634& -0.662741& 0.420455
\end{array}
\right)
\end{equation}
In the same units, the rotation mobility $\Mopo$ is 
\begin{equation}\label{eq:MopOmegaE1chargecube} 
\Mop_{\Omega E} =
\left(
\begin{array}{ccc}
 0.413811& -4.16162& 2.48241\\
 1.21999& 0.305149& -0.666799\\
-1.41969& 2.93046& -1.31055
\end{array}
\right)
\end{equation}
These matrices are precise for our stokeslet object approximating a cube; they are found by solution of large matrix equations with machine precision.  In view of the results Sec. \ref{sec:validation}, they should be a good approximation to the behavior of an actual charged cube, as well.

As anticipated in Sec. \ref{sec:pointcharge}, the matrix is singular; its determinant vanishes.  The quantitative characteristics of $\Mopo$ are shown graphically in Fig. \ref{fig:1ChargeCubeAxesTrace}.  For any initial state of the system, the position and orientation after a short time step $\Delta t$ are determined by $\vector V$ and $\vector \Omega$.  Using these, we may calculate the position and orientation at the end of the time step.  We may also calculate the new $\Mop_{VE}$ and $\Mopo$ by rotating the original $\Mop_{VE}$ and $\Mopo$ using the rotation matrix $R(\Delta t)\definedas \exp[\xop{\Omega} \Delta t]$ via Eq. \eqref{eq:MopRop}.  Repeated iteration of this procedure gives the matrices and the orientation after any number of time steps.

\begin{figure}[htbp]
\includegraphics[width=\columnwidth]{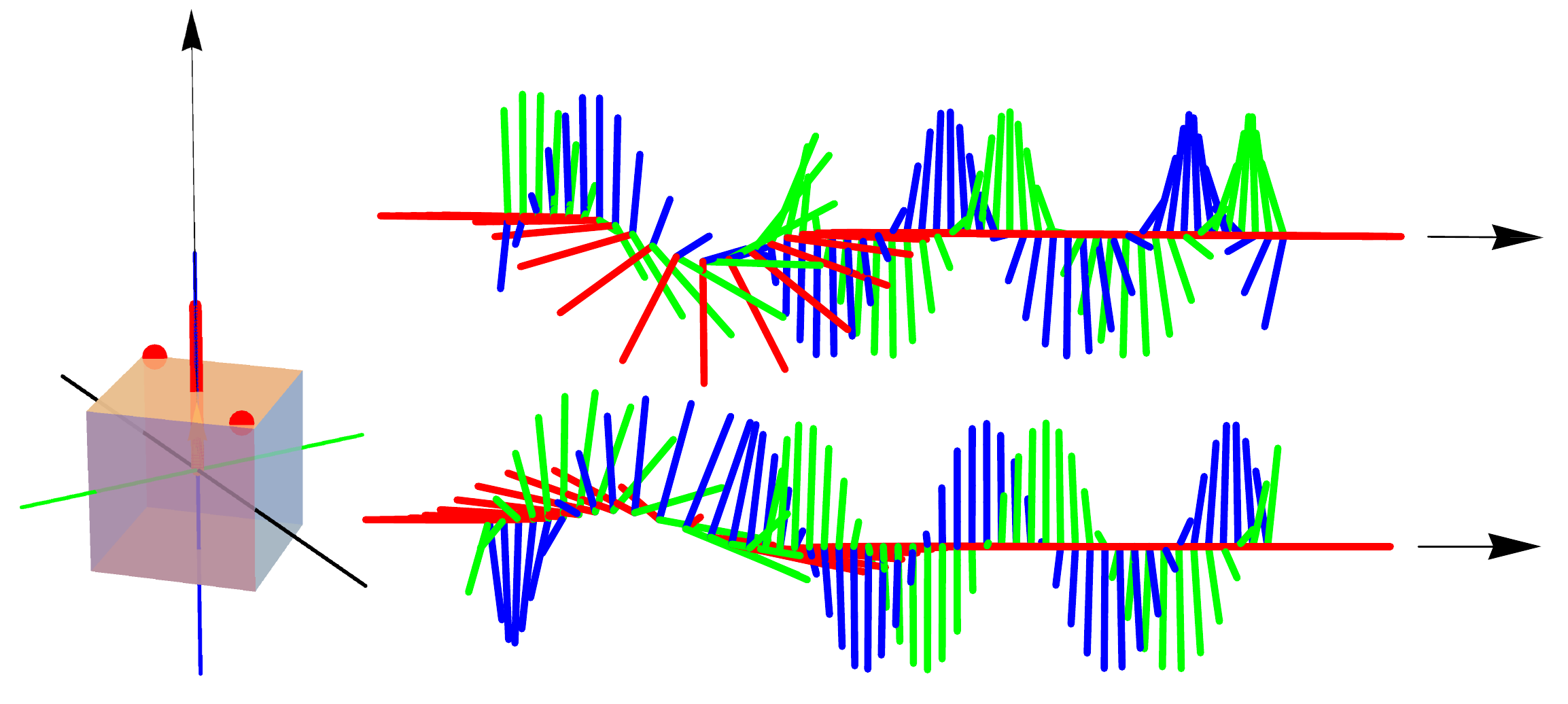}
\caption{\label{fig:rotationbyPi50-50weights} 
Motion of a symmetrical two-charge cube. top: 
the cube showing the charges and the characteristic axes defined in Fig. \ref{fig:1ChargeCubeAxesTrace}.
bottom: multiple-exposure view with the conventions of Fig. \ref{fig:1ChargeCubeAxesTrace}.  The cube shows a strong negative helicity, making about one turn for one cube-length of translation.
}
\end{figure}

\begin{figure}[htbp]
\includegraphics[width=\columnwidth]{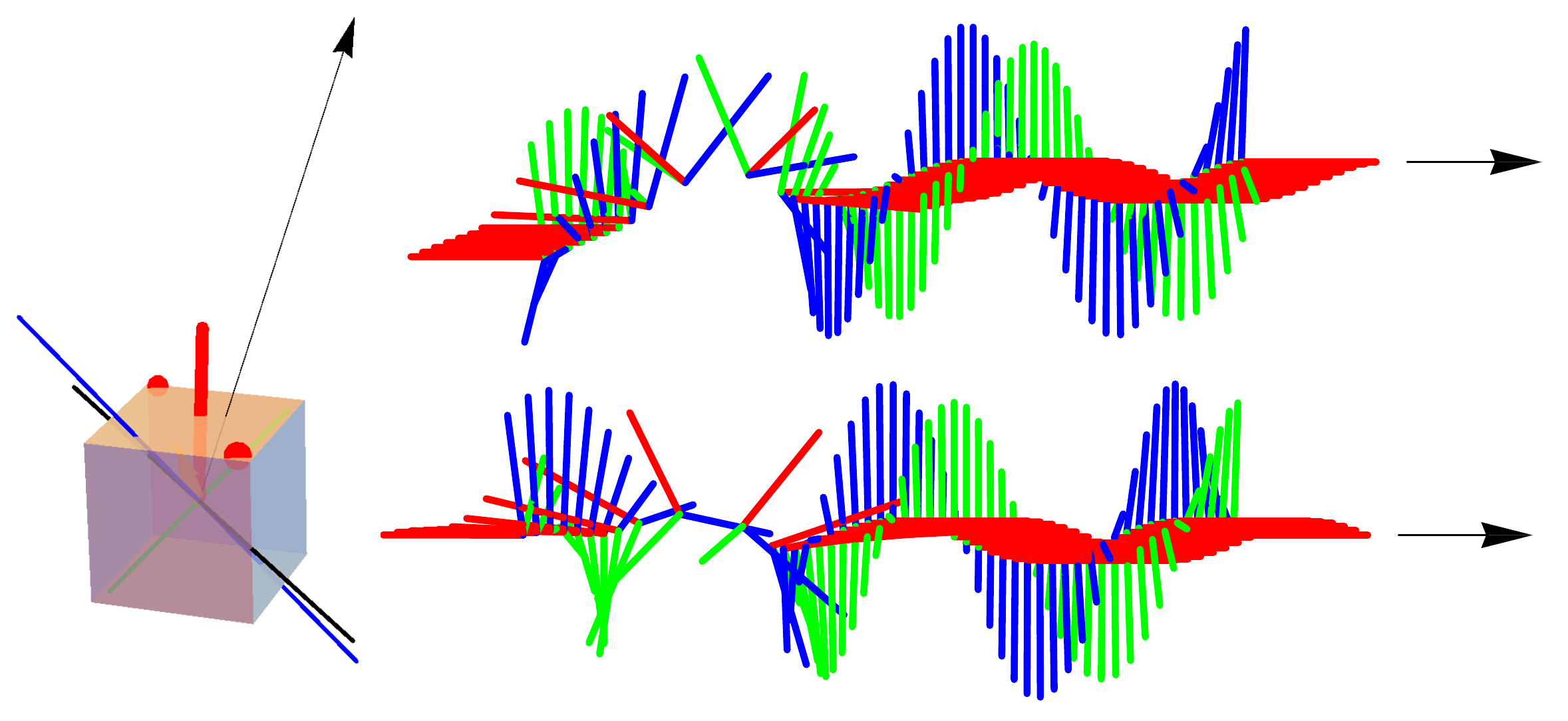}
\caption{\label{fig:rotationbyPi70-30weights} 
Motion of a symmetrical two-charge cube showing effect of unequal charges in 70:30 ratio. Left: 
the cube showing the larger charge near front edge and the characteristic axes defined in Fig. \ref{fig:1ChargeCubeAxesTrace}.  The aligning direction and the dipole are no longer perpendicular to the face.
Right: multiple-exposure view with the conventions of Fig. \ref{fig:rotationbyPi50-50weights}.  The motion is now helical}
\end{figure}

\begin{figure}[htbp]
\includegraphics[width=\columnwidth]{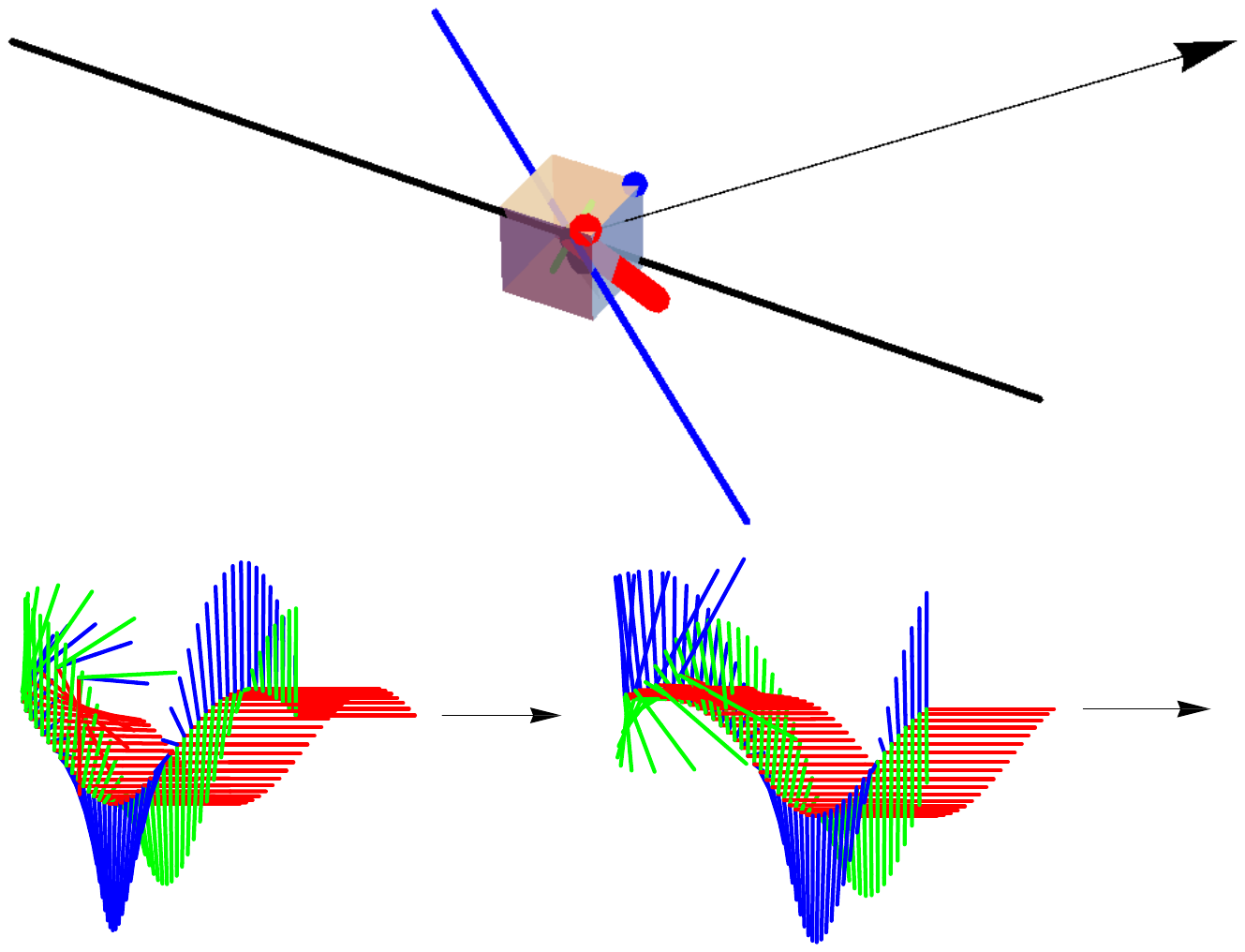}
\caption{\label{fig:rotationbyPiOver2-1-2weights} 
Motion of a cube with two charges on adjacent edges of a face. Charge ratio was -1:2.  The net charge is the same as in the figures above. Left: 
the cube showing smaller negative charge in darker color (blue).  The characteristic axes are defined in Fig. \ref{fig:1ChargeCubeAxesTrace}.  The strong differences from cases above results from the presence of opposite charges.  Right: multiple-exposure view with the conventions of Fig. \ref{fig:rotationbyPi50-50weights}.  Initial orientation was arbitrary.  Two orthogonal views are shown, as in previous figures.  Helical radius is a substantial fraction of the cube length.}
\end{figure}

The addition of a second point charge adds persistent chiral response.  We first show the symmetric case treated in Sec. \ref{sec:twopointcharges}.  In Fig. \ref{fig:rotationbyPi50-50weights} a second point charge is added to the cube of Fig. \ref{fig:1ChargeCubeAxesTrace}, at the symmetrically opposite  edge, maintaining a total charge of 1.  The  $\Mopo$ for this second charge is found by rotating the  $\Mopo$ of Eq. \eqref{eq:MopOmegaE1chargecube} by a half turn about the $\hat 3$ axis. The response matrix of the two-charge  object is the sum of the responses the two singly charged objects, each with charge 1/2 \cite{Note4}.
Thus by adding the rotated matrix to its unrotated counterpart, we obtain the $\Mopo$ matrix for the two-charge system.  The translation matrix $\Mop_{VE}$ is found analogously.

Much of the behavior can be anticipated by symmetry.  the alignment axis is evidently in the $\hat 3$ direction, and the velocity of the cube when oriented in this direction must also be along $\hat 3$.  There is a rotation since the two charges produce equal and nonzero torques about this axis.  

The chiral motion seen for this symmetric shape persists for general objects with asymmetric charge magnitudes and position, as shown in Figs. \ref{fig:rotationbyPi70-30weights}, \ref{fig:rotationbyPiOver2-1-2weights}.  Altering the charge ratio by a factor of order unity does not strongly degrade the chiral motion 
\begin{figure}[htbp]
\includegraphics[width=\columnwidth]{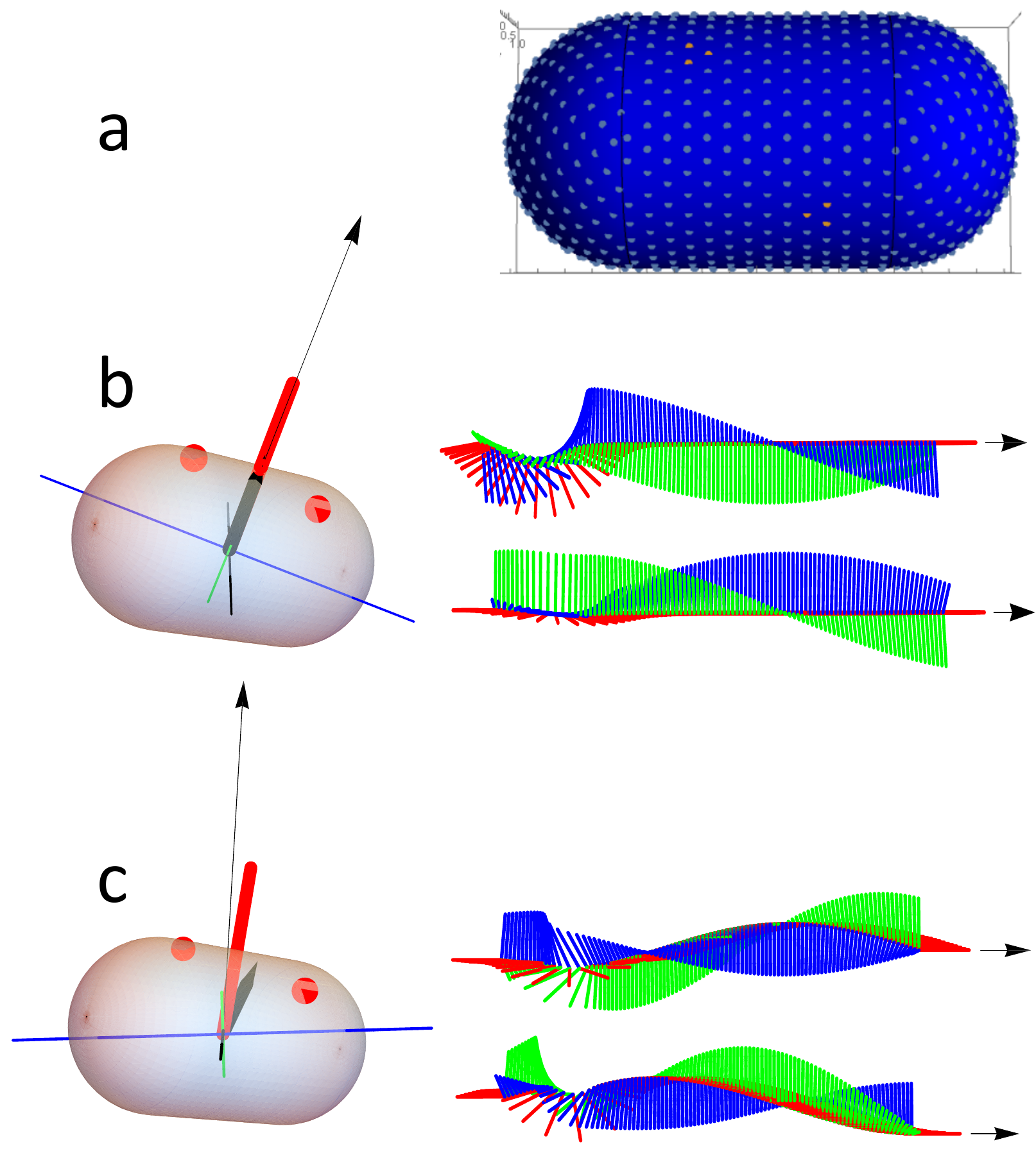}
\caption{\label{fig:PillTwoChargeSymmetric} 
Motion of capsule shapes with aspect ratio 2:1, bearing two localized charges  totaling 1.  a) stokeslet representation of the object showing the two charged regions as six light-colored dots.  
b) left: characteristic axes of the electrophoretic mobility using the conventions of Fig. \ref{fig:1ChargeCubeAxesTrace}. Short, heavy black arrow is equal to the asymptotic velocity. right: two orthogonal views of the motion using the conventions of Fig. \ref{fig:1ChargeCubeAxesTrace}. Basis vectors have unit length, equal to the cylinder diameter. Duration of the trajectory was time for the uniform cylinder to move 18 diameters.  After a rapid reorientation the capsule rotates slowly about its symmetry axis.  c) same as b, with charge ratio changed to 7:3.
}
\end{figure}

\subsubsection{Capsule shape}
The chiral motion reported above generalizes to other shapes.  We chose the spherocylinder or capsule shape mentioned in Sec. \ref{sec:validation} to show that the chiral motion occurs for smooth shapes as well as the sharp-edged cube shape.   We used the strategy suggested in Sec. \ref{sec:twopointcharges} to guess appropriate charge configurations.  As with the cube, we chose regions where the surface field had a component perpendicular to both the applied field and the displacement from the center for a given external field.  To gauge the effect of spreading the charge, we distributed the charge on three triangular clusters of stokeslets rather than on single stokeslets. With a single triangle of charge, the null eigenvalues in $\Mop_{VE}$ and $\Mopo$ were replaced by very small ones---$10^{-2}$ for $\Mop_{VE}$ and $10^{-4}$ for $\Mopo$.  We attribute these nonzero values to using a charge with a nonzero spatial extent.  We then placed a second charge at a symmetric point such that the capsule would have twofold symmetry about the transverse axis, as shown in Figure \ref{fig:PillTwoChargeSymmetric}b.  In Figure \ref{fig:PillTwoChargeSymmetric}c we show the effect of unequal charges.

\subsubsection{Summary}
The above examples indicate that strongly chiral motion is readily attainable with a wide range of simple configurations.  We summarize the quantitative features of the chiral motion in Table \ref{tab:specs}.  These objects generally rotate about their stable axis at a rate of order unity when scaled by the smoluchowski speed of the object and the object's size.  We could change the placement and relative magnitudes over significant ranges without strongly degrading the chiral motion.   The constraints on shapes needed for chiral motion were modest.  Even the high-order anisotropy of a cube is sufficient.  Thus many colloidal particles encountered in nature should show a distinctive and observable chiral signature.  We discuss this prospect in the next section.

\begin{table}[htp] 
\caption{\label{tab:specs} Chiral features of objects with two localized charges.  Unit of length is side length for cube, diameter for capsule. ``Position symmetry" indicates whether the second charge was placed in a symmetric position so that the rotation axis could be inferred by symmetry.  $\Omega$ is in units of speed of the uniformly charged object per unit length. Positive $\Omega$ indicates right-hand rotation. Pitch is number of lengths moved in one rotation.  Radius is radius of helical path of the middle of the object.\medskip}
\begin{ruledtabular}
\begin{tabular} { c  c  c  c  c  c  c  }
object~ & Fig.~ & charge~ & position~ & $\Omega$~ & pitch~ & radius \\
 & & ratio & symmetry~ & & & \\
& & & & & &\\
cube & \ref{fig:rotationbyPi50-50weights}     & 1:1  & yes &-1.3 &2.0 &0 \\
     & \ref{fig:rotationbyPi70-30weights}     & 7:3  & yes &-1.0 &2.1 & 0.14 \\
     & \ref{fig:rotationbyPiOver2-1-2weights} & -1:2  & no & -1.4& 2.3 & 0.46\\
capsule & \ref{fig:PillTwoChargeSymmetric}b   & 1:1  & yes & +0.16 & 22 & 0.02\\
        & \ref{fig:PillTwoChargeSymmetric}c   & 7:3  & yes & +0.25 & 11 & 0.55\\
\end{tabular}
\end{ruledtabular}
\end{table}%

\section{Discussion}\label{sec:discussion}

Our goal in this study was to provide convincing evidence of experimentally accessible chiral signatures in colloidal particles.  To this end we developed a concrete numerical method capable of giving reliable estimates of the rate of chiral rotation for given shapes. The chiral effects were of order unity on the scale of conventional electrophoretic motion.  Thus apparatus that can track conventional electrophoretic phenomena should be able to track these motions.   

Here we discuss limitations of our results for predicting measurements.  We discuss how our point-charge results can be used to estimate more realistic cases of distributed charge.  We note how this chiral response can be used to organize a suspension of like bodies.  We comment on the hydrodynamic interactions expected between such bodies.  Finally we survey the implications of our findings beyond electrophoresis.  .  

\subsection{Limitations}\label{sec:limitations}

At first sight our stokeslet-object representation appears as a major limitation in accuracy, particularly when contrasted with boundary element methods such as \cite{Youngren:1975rm}.   The boundary element method views the body as a polyhedron and generates the flow by matching hydrodynamic boundary conditions at the center of each face.  In contrast, we represent the body as a dilute set of stokes spheres distributed over the body surface.  There is no explicit representation of boundary conditions on a solid surface.  Thus flow on the outside can penetrate into the interior.  Nevertheless, these stokeslets give a good representation of flow around a solid body, thanks to hydrodynamic screening, as explained in the Appendix.  

As seen above, the limited resolution of our mesh limits the precision of our predictions to the 5-10 percent level. The worst discrepancies appear to result from sharp features like the cube edges.  Increasing the number of stokeslets improves the agreement, as with Ref. \cite{Mowitz:2017kx}.  The calculations reported here were feasible on a personal computer.  But the computation time increases rapidly with number of stokeslets, so that this method would be inefficient for precise computation.  

Beyond this, the assumptions used in our calculations set further limits on experimental predictability.  Most notably, our theory is confined to linear responses to the external field $E^0$.  Thus it takes no account of effects like dielectrophoresis (quadratic in $E^0$) or dependence of the screening charge distribution on $E^0$.  This restriction and experimental limitations limit experimental values of $E^0$ to the order of 100 V/cm. Further, the realities of ionic equilibrium in typical solutions limit the attainable zeta potentials to tens of millivolts or less.  

The examples above show chiral motions of point-charged bodies comparable to uniformly charged bodies with the same total charge.  But this comparison can be misleading.  A given, attainable zeta potential confined to a small fraction of the surface necessarily means a small Smoluchowski velocity, proportional to the relative area of the charged region.  Thus the point charges discussed here constitute an impractical limit.  Instead, one must inevitably consider charge spread over some minimal fraction of the surface.  This spreading of the charge over a finite area necessarily diminishes the chiral effects.  Indeed, if the charge is spread uniformly over the surface, all chiral effects must cease: only the scalar response dictated by the Morrison theorem \cite{Morrison} is possible. Thus, as one expands from zero the region on the body where the zeta potential is nonzero, the chiral effects (such as $\Omega$) first increase in proportion to the total charge, but then decrease as the charged area becomes comparable to the total area.

There are also practical limitations on the range of body sizes that can show significant chiral motion.  For a given fixed zeta potential distribution and a fixed Smoluchowski speed, $\Omega$ varies inversely with the size of the body.  Small bodies rotate faster, but they also undergo faster rotational diffusion.  For the typical conditions envisaged above, rotational diffusion swamps chiral rotation for bodies smaller than the scale of 100 nm.  The maximum size is set by experimental convenience.  Smoluchowski speeds are typically on the order of 1 mm/sec or slower, $\Omega$ becomes inconveniently slow for particles much larger than a millimeter.  

Not all charge distributions are expected to give the kind of axial alignment shown here. This alignment requires that $\Mopo$ have two nonreal eigenvalues \cite{Gonzalez04}.  However, the $\Mopo$ matrix may have a full set of simple, real eigenvectors.  Then there is more than one locally stable orientation, as shown in Fig. \ref{fig:Orbits}b.  While this sort of motion is less predictable and reliable than the aligning cases shown above, the richness of possible behaviors is greater.  Such motion, especially in time dependent external fields, potentially gives further means for organizing colloidal dispersions along the lines discussed below.  

\subsection{Uses}\label{sec:uses}
Conventional electrophoresis is widely used to separate and characterize objects of molecular or colloidal scale according to size and charge\cite{Delgado:2007qy}.  The bodies studied here have a richer response: their motion depends on the 18 parameters of the $\Mop_{VE}$ and $\Mopo$ tensors.  These can all be measured in principle by varying the field direction and observing the resulting motion.  This gives a means of distinguishing many aspects of the shape and the charge distribution of a body such as a cell or virus.  Each element of $\Mop_{VE}$ and $\Mopo$ determines a particular moment of the zeta potential over the surface.  For spheres, these are combinations of conventional monopole, dipole and quadrupole moments\cite{AndersonSpheres}.  For general shapes the moment functions corresponding to the $\Mop_{VE}$ and $\Mopo$ elements are specific to the shape, as found by Teubner\cite{Teubner:1982kq}.  Thus measuring the tensorial response, both chiral and nonchiral, can in principle provide a substantial insight into the charge distribution over a body.  Since this distribution often depends on conditions such as pH, a ready means of observing these changes is certainly desirable, and electrophoresis offers such a means.  

Even when the shape of the body and the corresponding Teubner moments are known, the utility of the electrophoretic measurement in constraining the charge distribution hinges on the distinguishability of the these moments.  Thus knowing these moments for a given shape is important.  The method of Sec. \ref{sec:implementation} gives partial information by predicting the motion expected from an assumed charge distribution.  In a recent work we showed how to extend these methods to obtain the Teubner moments explicitly for a given stokeslet object\cite{Witten:2019fv}, thus providing a direct transform from charge distribution to the corresponding $\Mop_{VE}$ and $\Mopo$.  

A further use of tensorial electrophoresis is to drive co-ordinated motion using time-dependent fields.  In conventional electrophoresis, the motion simply follows the instantaneous field.  However the objects studied above undergo a well-defined transient motion whenever $E^0$ is changed.  This transient response provides a handle for manipulating the orientation of the object.  When applied to a dispersion of many like objects in random orientations, a uniform time-dependent $\vector E^0$ can bring them all into the same orientation and synchronous rotation\cite{Eaton-Moths-Witten,Moths-Witten2}.

\subsection{Interactions}\label{sec:interactions}
The treatment above considers a single body in isolation.  But most of the uses contemplated above involve suspensions of many particles in a common fluid.  Thus the flow caused by one body's motion must influence nearby bodies.  These hydrodynamic interactions produce strong collective effects when the flow is caused by external body forces, as in gravitational sedimentation.  Then the far-field velocity is a force monopole with a $1/r$ dependence on separation $r$. When the sedimenting bodies are asymmetric, they perturb each other's orientation and collective motion substantially\cite{Goldfriend:2017mz,TomerHaim1,TomerHaim2}. Generally these interactions arise from the gradient of the velocity caused by the driven particle.  In electrophoresis with no net force on the body, the asymptotic field falls off at least as fast as $1/r^2$---a force dipole.  This force dipole flow occurs \eg when a charged sphere drags a neutral sphere tethered to it. Its gradient thus falls off as $1/r^3$, so that hydrodynamic interaction is predominantly local. \footnote{
An additional form of interaction is electrostatic. The $\vector E^s$ of one body is a response to the local $\vector E_{01}$ at that body.  This local $\vector E_{01}$ is the external field $\vector E_0$ plus the field induced on the second body, $\vector E_p$.  At long distance $r$ from the second body, $\vector E_p$ is a dipole field, of order $E_0/r^3$.  Thus the first body experiences an external field altered by a factor of of order $E_p/E_0 \goesas 1/r^3$. This electrostatic effect falls off comparably to the hydrodynamic effect treated in the main text.  We note that effect is independent of any explicit charge on the bodies, so it must dominate if the explicit charge is very weak.  
} 
Still, the short range interactions due to electrophoretic driving are a novel potential source of collective effects, especially collective chiral effects\cite{Bricard:2013pr,Nash:2015sh}.

\subsection{Generalization}\label{sec:generalization}
Electrophoresis generates motion without injecting momentum into the host fluid.  This distinctive feature is shared by many forms of phoresis.  Further, many of these phoresis effects arise from a thin shear layer resulting from the phoretic driving, \eg from chemical or thermal nonequilibrium.  In these respects, they may be treated in parallel with electrophoresis\cite{Anderson1989colloid}.  As with electrophoresis, these motions are proportional to an externally imposed gradient.  Both the difficulties and the interest of studying asymmetrical shapes and nonuniform local shear are parallel to the electrophoretic case.  Thus most of the methodology used here and many of the phenomena predicted have should also have parallels for general forms of phoresis.  In particular the flexibility of the stokeslet object approach is equally advantageous for studying these phenomena.

Further forms of driven motion, while caused by a thin shear layer, do not arise from an external gradient.  Instead they arise from some autonomous process intrinsic to the body, such as beating cilia on the surface of a living organism\cite{Marchetti:2013pi} or a fluid instability caused by a chemical reaction\cite{Walther:2013jt, Maass:2016sy}.  Here too the interest in creating rotation in asymmetric bodies is strong.  Thus the stokeslet object methodology is of potential value.

\section{Conclusion} \label{sec:Conclusion}
Experimental study of tensorial electrophoresis has been meager up to now, despite its fundamental nature and its intriquing consequences.  Above we provided a general scheme for predicting these consequences from well-defined properties of the moving body.  The scheme is suitable for anticipating the motion given the body's properties without the restrictions of previous methods.  We hope these predictions will stimulate experiments to eclipse or disprove them. 

    \section*{Acknowledgments}     
 The authors are grateful to Haim Diamant and Tomer Goldfriend for continual advice and insights throughout this work. Tomer Goldfriend's insightful reading of the manuscript led to many improvements. We thank Gerwin Koolstra for his calculation of the electrical potential contours shown in Fig. \ref{fig:CubeWithEs}.
We thank Ilona Kreszchmar for consultations about experiments and for inspiring the choice of the capsule shape. This work was partially supported by the University of Chicago Materials Research Science and Engineering Center, which is funded by the National Science Foundation under award number DMR-1420709.


\section*{Appendix: Electric field and drag calculations for stokeslet objects}
\subsection{determination of surface field $\vector E^s$}
The determinating $\vector E^s$ is a standard problem in numerical electrostatics, and many software packages exist to solve it\cite{ANSYSMaxwell}.  We use the method below for the sake of consistent methodology.

The surface field $\vector E^s_i$ is the sum of the imposed field $\vector E^0$ and the induced field at $i$ caused by polarization charges $Q_j$ at each mesh point $j$.  This surface charge is present because the current density $J^s$ and hence $E^s$ must have no normal component on this insulating body.  The contribution $\vector E_{ij}$ at point $i$ due to charge $j\notequal i$ is given by Coulomb's law: 
\begin{equation}
\vector E_{ij} = {Q_j \over 4\pi\epsilon_0} ~{(\vector r_i - \vector r_j) \over |\vector r_i - \vector r_j|^3}.
\end{equation}
There is also a contribution to $\vector E^s_i$ due to $Q_i$ itself.  Here the approximation of a point charge is inadequate; thus we represent $Q_i$ as a uniform disk of charge whose area is the Voronoi area $A_i$ of the mesh point $i$.  
The induced field at $i$ due to $Q_i$ is then $Q_i ~\hat n_i/(\epsilon_0 A_i)$.
The net field $\vector E^s_i$ at point $i$ is then
\begin{equation}\label{surfaceField}
\vector E^s_i = \vector E^0 + \sum_{j\notequal i} {Q_j \over 4\pi\epsilon_0} ~{(\vector r_i - \vector r_j) \over |\vector r_i - \vector r_j|^3}
+ Q_i \hat n_i/(\epsilon_0 A_i)
\end{equation}
We now require that the normal component $\vector E^s_i \cdot \hat n_i$ vanish for every $i$.  There is one such requirement for each of the charges $Q_i$. Thus Eqs. \eqref{surfaceField} determine the $Q_i$'s for any given imposed field $\vector E^0$.  Using these $Q_i$'s in Eq. \eqref{surfaceField} gives the resulting tangential $\vector E^s_i$ field.  We determine these $\vector E^s_i$ for each of three orthogonal directions of $\vector E^0$.  This allows us to infer the $\vector E^s_i$ for any $\vector E^0$ by superposition.  

\subsection{determining sedimentation drag coefficients}
We consider the stokeslets to be moving as a rigid body with velocity $\vector V$ and angular velocity $\vector \Omega$ about some given origin.  Thus stokeslet $i$ moves at a velocity  $\vector u_i$ given by 
\begin{equation}\label{eq:ui}
\vector u_i  = V + \vector \Omega \cross \vector r_i .
\end{equation}
The fluid at $i$ has a velocity $v_i$ arising from the drag forces on the other stokeslets, denoted $\vector g_j$.  Each stokeslet $j$ contributes a velocity at $i$ denoted $\vector v_{ij}$ and given by
\begin{equation}
\vector v^s_{ij} = {1 \over 8 \pi \eta}{\vector g_j + (\vector g_j \cdot \hat r) ~ \hat r \over |r|},
\end{equation}
In general the fluid velocity $\vector v_i$ is not equal to the stokeslet's imposed velocity $\vector u_i$; this difference in velocity entails a drag force $\vector g_i$ on the fluid given by $\vector g_i = \gamma (\vector u_i - \vector v_i)$. 
Using Eq. \eqref{eq:ui} for $\vector u_i$ and $\sum_j \vector v^s_{ij}$ for $\vector v_i$ gives a closed set of linear equations for the forces $\vector g_i$.
The total force $\vector F$ on the fluid is then $\sum_i \vector g_i$; the total torque $\vector \tau$ is $\sum_i \vector r_i \cross \vector g_i$.  This force is proportional to the imposed $\vector V$ and to $\vector \Omega$.  The proportionality of $\vector F$ to $\vector V$ defines the translational resistance matrix ${\Kop}$ of Happel and Brenner\cite{Happel-Brenner} Ch. 5.  In general the proportionalities of $\vector F$ and $\vector \tau$ to $\vector V$ and $\vector \Omega$ define the full set of four resistance matrices.  In combination these give the $(\vector F, \vector \tau)$ for given $(\vector V, \vector  \Omega)$ via the $6\cross 6$ resistance tensor ${\cal R}$ such that $(\vector F, \vector \tau) = {\cal R} (\vector V, \vector \Omega)$.  
For use below, we invert  this ${\cal R}$ matrix to find the proportionality of $\vector V$ to $\vector F$ and $\vector \tau$.  This defines the Stokes mobility matrices  $\Mop_{VF}$ and $\Mop_{V \tau}$.  The matrices giving the proportionality of $\vector \Omega$ to $\vector F$ and $\vector \tau$ are defined similarly:
\begin{eqnarray}
\vector V =& ~\Mop_{V F} \cdot \vector F + \Mop_{V \tau}\cdot  \vector \tau \\
{\rm and} &\nonumber\\
\vector \Omega =& ~\Mop_{\Omega F}\cdot \vector F + \Mop_{\Omega\tau } \cdot \vector \tau 
\end{eqnarray}

Sufficiently many stokeslets produce strong screening so that the interior fluid moves with the body.  The area fraction $\phi_a$ of stokes spheres of radius $a$ needed to produce this strong screening in an object of size $R$ is of order $(a/R)$.  Thus the spheres can be arbitrarily dilute if they are sufficiently small and numerous\cite{Mowitz:2017kx}.  
\subsection{determining the electrophoretic forces $\vector f_i$}
The imposed velocities $\vector v^s_i$ are linearly related to the $\vector f_i$ using Eq. \eqref{eq:Oseen}:
\begin{equation}
\vector v^s_i = \sum_j \vector v^s_{ij} = \\
{1 \over 8 \pi \eta}\sum_j {\vector f_j + (\vector f_j \cdot \hat r) ~ \hat r \over |r|},
\end{equation}

As with the surface fields $\vector E^s_i$ the term $\vector v^s_{ii}$, \ie the velocity at $i$ due to the stokeslet at $i$ itself requires special treatment.  A point stokeslet at $i$ would give a divergent velocity at $i$.   To remove this unphysical divergence, we replace the point with uniformly distributed force distributed over a disk with the Voronoi area $A_i$.  The resulting $\vector v^s_{ii}$ is the integral of this force at the center of the disk.  Combining the contribution from $\vector r$ with that at $-\vector r$, we see that only the first term in Eq. \eqref{eq:Oseen} contributes, so that $\vector v^s_{ii}$ points along $\vector f_i$.  Specifically, 
\begin{equation}\label{selfVelocity}
\vector v^s_{ii} = {|f_i| \over 8\eta \sqrt{\pi A_i}}~ [3 \hat f_i - (\hat n_i \cdot \hat f_i)~ \hat n_i]
\end{equation} 


\bibliography{StokesletElectrophoresis}

\end{document}